%% file: microfluidics_gess_arxiv_v2.tex
 \documentclass{scrartcl}
 \usepackage{epsfig}
 \input{File_Setup.tex}
 \usepackage{amsmath,amssymb,mathpazo,fullpage,microtype}
\usepackage{cleveref}

 \graphicspath{{Graphics/}}

 \definecolor{webgreen}{rgb}{0,.35,0}
 \definecolor{webbrown}{rgb}{.6,0,0}
 \definecolor{RoyalBlue}{rgb}{0,0,0.9}
 
 \hypersetup{
    colorlinks=true, linktocpage=true, pdfstartpage=1, pdfstartview=FitV,
    breaklinks=true, pdfpagemode=UseNone, pageanchor=true, pdfpagemode=UseOutlines,
    plainpages=false, bookmarksnumbered, bookmarksopen=true, bookmarksopenlevel=1,
    hypertexnames=true, pdfhighlight=/O,
    urlcolor=webbrown, linkcolor=RoyalBlue, citecolor=webgreen,
    pdfauthor={},
    pdfkeywords={},
    pdfcreator={XeLaTeX},
    pdfproducer={LaTeX with hyperref}
 }

\newcommand{\ud}{\mathrm{d}}

\title{\huge\bfseries{Multiscale Model of Clogging in Microfluidic Devices with Grid-like Geometries}}

 \date{}
 \usepackage{authblk}
 \author[$\star$]{Gess Kelly}
 \author[$\dagger$]{Thomas G. Fai}
 \affil[$\star$]{{\small Martin A. Fisher School of Physics, Brandeis University, Waltham, MA 02453}}
 \affil[$\dagger$]{{\small Department of Mathematics and Volen Center for Complex Systems, Brandeis University, Waltham, MA 02453}}

\begin{document}

%\begin{fmtext}

\maketitle
\begin{abstract}
We propose a coarse-grained theoretical model to capture the aging of microfluidic devices under different conditions including constant applied flow rate and constant applied pressure gradient. Microfluidic devices that sort cells by their deformability hold significant promise for medical applications. However, clogging in these microfluidic systems causes their properties to change over time and potentially limits their reliability. We compare the results of the coarse-grained model to those of stochastic simulations and to existing theoretical studies. Lastly, we apply the model to experimental data on the clogging of sickle red blood cells and discuss its wider applicability.
\end{abstract}

\pagestyle{plain}

\section{Introduction}
Microfluidic devices play an important role in various disciplines, with applications ranging from inkjet printing, cooling of integrated circuits, mimicking porous structures in studies of soil improvement, and biomedical instruments \cite{icircuit1, SuWenjing2016FimA, RappBastianE2016MMMa,soil1,soil2}. In some cases, clogging hinders the performance of these devices or decreases their efficiency in transporting particles. For instance, in microfluidic cell sorters that sort sickle red blood cells based on their deformability, clogging leads to greater unpredictability in the device behavior. Previous studies have shown that clogging contributes to a faster fouling of the device \cite{canc_sort2, culture_fouling, cl_fouling, softmatter, canc_sort}.  Investigating the detailed nature of clogging in these microfluidic devices provides an opportunity to gain a deeper understanding of their properties and how their performance can be improved. \par
More specifically, we focus on clogging in diagnostic devices in which microchannels are used to sort cells based on their deformability.
%\end{fmtext}

%%%%%%%%%%%%%%% End of first page %%%%%%%%%%%%%%%%%%%%%

\noindent Diseases such as sickle cell disease, malaria, and some types of cancer can be detected by microfluidic devices that detect, sort, or capture unhealthy cells based on changes in their physical properties \cite{Shelbymalaria_microfluidic, cancer1, Du1422, WuTenghu2013malaria, malaria2}. {These devices are unified in their design by the principle that healthy deformable cells squeeze through narrow microchannels more easily than rigid cells.} {For instance, devices that sort cells infected by malaria can help in assessing the stage of the disease by finding the critical pressure required for squeezing the cells in a channel with a specific geometry \cite{WuTenghu2013malaria}.  In other contexts, the geometry chosen for the microchannels is optimized to separate infected cells from the general cell population in a diluted sample of blood \cite{BowHansen2011malaria}. Another example is devices that can estimate the severity of sickle cell disease by monitoring the fraction of clogged channels or an index based on this fraction as cells flow through the channels and block the channels \cite{ManYuncheng2020OI, Du1422}.}

{Many questions regarding the behavior of these devices remain unanswered. These questions include} how clogging changes the device properties over time and how clogging depends on the geometry of the device. Although here we focus on biomedical devices, we remark that the issue of clogging  pertains to a wide range of applications including water filtration, chemical extraction, soil rehabilitation using bacteria, and traffic congestion \cite{grav, ped, ped_clg}. \par

Microfluidic cell sorters come in a variety of geometries. In this paper, we focus on clogging in systems that have grid-like geometries consisting of identical rows and columns. Moreover, here we consider channels with uniform size of the same order of magnitude as the particles. This further narrows down the problem to the cases where particles block the channels by the sieving mechanism \cite{SauretAlban2014Cbsi,dis_sorrel} or complete blocking \cite{sanaei2019}. \par 
{Our consideration of grid-like structures may be viewed as an extension to the geometry previously studied in works such as \cite{talbot_model,talbot_barre_reversible} and \cite{SauretAlban2018Goci} where Sauret et al. (2018) follow similar calculations presented by Talbot and Barr\'e (2015) and confirm their results by experiments. We apply our model to this simpler geometry of parallel channels and show that it agrees with these previous works.} Although here we focus on grid-like geometries for simplicity and ease of analysis, we note that a variety of geometries have been studied previously and that clogging may occur in various types of geometries. Experimental studies that consider a wider selection of geometries report that tortuosity exacerbates clogging \cite{connectivity}. Geometries with ladder-like connectivity \cite{droplet} are also common among microfluidic systems such as blood circuits in the zebrafish \cite{zebrafish}. In addition to connectivity and tortuosity, the ratio of pore size to particle size plays an important role in influencing the progress of clogging \cite{wyss}. \par
In our work, we develop a model in which channel clogging occurs instantaneously {\cite{SauretAlban2014Cbsi}}. This is a reasonable assumption in applications such as cell sorting, in which a single cell may become stuck in a channel and prevent other cells from passing. We acknowledge that this assumption {does not pertain to some} other contexts, such as the gradual aggregation of small particles that progressively occlude and eventually clog a channel. For simplicity, we assume laminar Poiseuille flow through each channel and invoke the electronic-hydraulic analogy \cite{Bruus, Stone2007}. This allows the application of elementary laws such as Ohm's and Kirchhoff's law \cite{AjdariArmand2004Sfin} following previous authors \cite{droplet, alim}.   \par 

To quantify and predict the dynamic behavior of microfluidic devices, here we develop a mathematical model for clogging using a multiscale approach that combines mean-field models, stochastic simulations, and analytical approximations. We study various specific cases that may be generalized to other geometries and clogging rate models. Within our model, the dynamics of clogging in a device are captured in terms of a specified clogging rate function. We explore the model-space of clogging rate functions under various conditions including constant total flow rate, constant pressure difference across the device, and independent channels. We encounter different behaviors, such as clogging in finite time, depending on the type of feedback in the system and the boundary conditions. We show that the results obtained from a simple mean-field model agree with those of previous probabilistic studies  on related problems \cite{talbot_model}. \par
Finally, we demonstrate an application of the model to {previous experiments} from \cite{Du1422} featuring a biomedical microfluidic device that sorts sickle red blood cells based on their deformability. In this device, all channels are arranged in a grid-like geometry, and their size has the same order of magnitude as the size of the cells. As the partial pressure of oxygen in this device is lowered, the sickled cells become more rigid, and they can no longer squeeze through the channels. We apply the mean-field model to predict the progression of the mean fraction of clogged channels over time. In this example, the clogging rate function reflects the changes in the deformability of the cells and how that affects their probability of clogging.
\section{Mathematical Modeling}\label{sec:model}
\begin{figure}[h]
    \centering
    \includegraphics[width = 12cm]{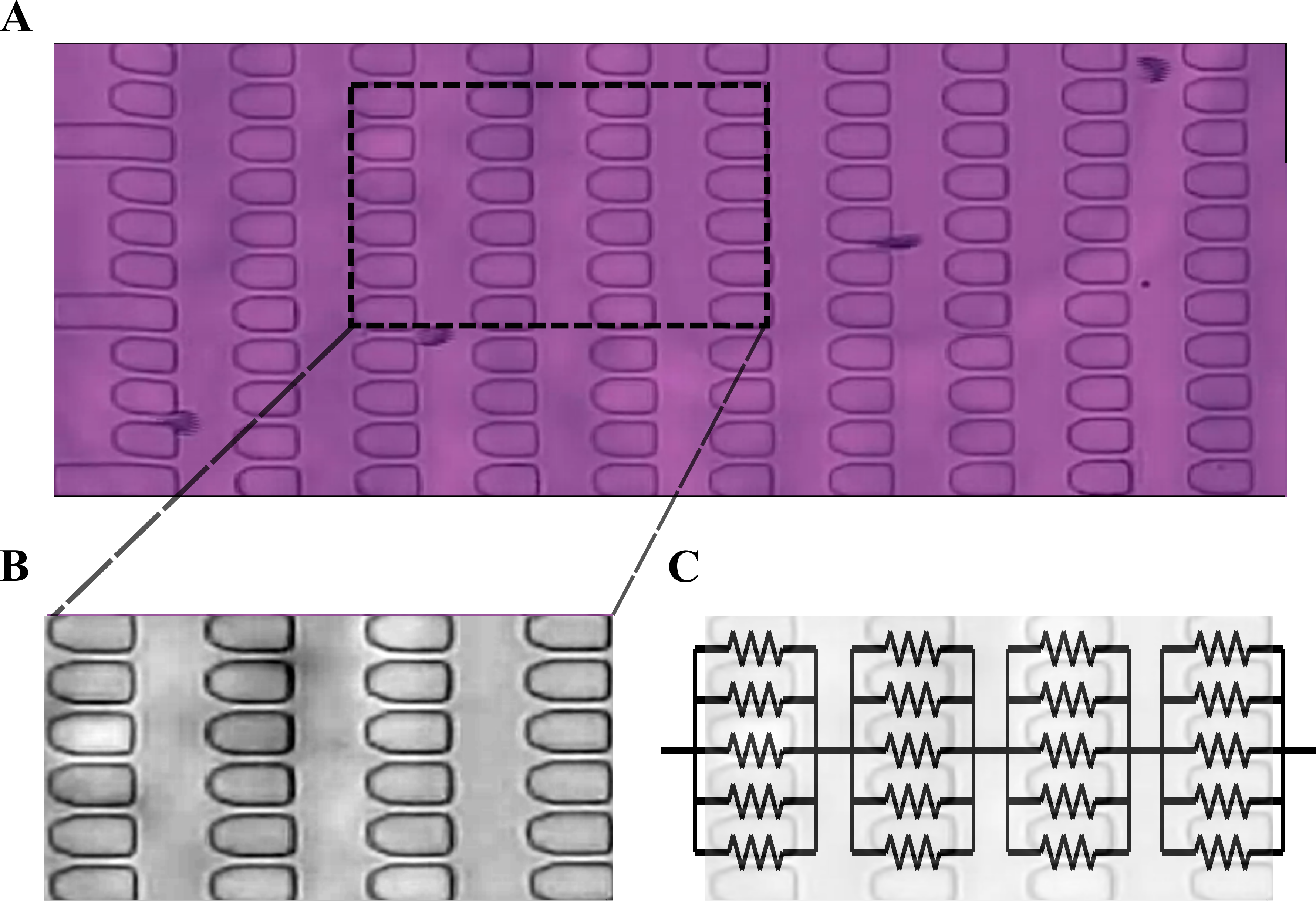}
    \caption{The hydraulic analogy allows us to represent the microfluidic device as a system of resistors. \textbf{A} Snapshot of a microfluidic device reproduced from \cite{Du1422} with permission of the publisher. A pressure gradient $\Delta P$ drives the flow from left to right and a sample of red blood cells squeeze through the channels.
\textbf{B}, \textbf{C}: An example of using hydraulic analogy to represent a section of the device, with {n = 4} columns and {m = 5} rows, by a resistor system consisting of {n = 4} sets of {m = 5} parallel resistors in series.}
    \label{fig:schem}
\end{figure}

Consider a device with identical microchannels organized in a grid-like geometry composed of $m$ rows and $n$ columns. We develop a mean-field model to study how the dynamics of the fraction of clogged channels $f(t)$ and clogging time $T_\text{clog}$,{ the time at which all channels in at least one column have clogged,} depend on the geometry and physical properties of the device. {For a more general discussion of mean-field modeling approaches we refer the reader to \cite{AldousDavidJ1999DaSM}}. We verify our model and test its limitations by comparing its predictions to the results of stochastic simulations and results from probability theory. Our model focuses on a grid-like geometry composed of rows and columns (see Figure \ref{fig:schem}). This geometry is similar to the one used in previous research on clogging in microfluidic channels in parallel rows (or bundles) \cite{talbot_model, SauretAlban2018Goci}. The clogging of individual channels is modeled by prescribing the probability that an open channel becomes instantaneously clogged by a cell. This generalizes the model of \cite{talbot_model} and \cite{SauretAlban2018Goci}, in which only a single column of channels was considered, to the case of an arbitrary number of columns connected in series.\par
 
We assume that the pressure difference across the device $\Delta P$ satisfies
\begin{equation}
 \Delta P = Q R,
    \label{PRQ}
\end{equation}
where $Q$ is the total flow rate and $R$ the hydraulic resistance, a function of the resistance of each channel, $r$, which depends on its shape and size and the fluid properties. Assuming laminar flow, which is reasonable for the microfluidic flows that we consider,  we use the Hagen-Poiseuille equation \cite{RappBastianE2016MMMa}. For instance, for a cylindrical channel with diameter $d$ and length $l$, 
\begin{equation}
    r = \frac{128 \eta l}{ \pi d^4},
\end{equation} 
where $\eta$ is {the dynamic viscosity of the fluid}. The hydraulic analogy {\cite{Bruus}} allows us to represent the channels in the same column as resistors in parallel, and columns of channels as resistor sets in series, and to use Kirchhoff's laws in order to compute the overall resistance $R$ in terms of the channel resistances $r$ and device connectivity. As an example, Figure \ref{fig:schem}\textbf{B} shows a subset of channels in a microfluidic sorting device, and Figure \ref{fig:schem}\textbf{C} shows a representation of the same system as a system of resistors. As an example, in an $m$ by $n$ device, the initial device resistance $R$ is given by
\begin{equation}
    R = n r/m,
\end{equation} according to Ohm's law. 
The system fails when all the channels in {at least} one of the columns of the device clog, at which point the fluid no longer flows through the device. As channels become clogged, the overall resistance $R$ increases and can be recalculated by applying Ohm's law once again. \par
Although the hydraulic analogy provides an efficient way of treating microfluidic structures, it has a number of limitations. For instance, it fails to encompass some of the details of the hydrodynamics such as spatial correlations between the channels.  In addition, while using the hydraulic analogy, we ignore the effects of individual particle projectiles and particle-particle interactions. While these effects may be negligible in the present case as we take clogging of each channel to be an instantaneous process, in a more general treatment of clogging in microfluidics, they could, in principle, influence the final results.  \par
To coarse-grain the clogging events of individual channels, we introduce a mean clogging rate function $\lambda(q, r)$ that depends on the channel flow rate $q$ and the channel resistance $r$. We assume that a particle clogs a channel with probability per time $\lambda(q, r)$. We model these clogging events as Poisson processes in which the channel clogging rate remains constant with respect to time between clogging events {\cite{SauretAlban2014Cbsi}}. Since we consider specific geometries in which all the channels are identical with the same resistance, we make the $r$-dependence implicit and focus on clogging rate functions that only depend on $q$. \par
Given these assumptions and taking the limit of a system with a large number of channels $N_0$, we use a mean-field model in which the number of clogged channels $N$ follows  
\begin{equation}
  \Dot{N}(t) = \lambda(q) (N_0-N(t)); \hspace{1cm} N(0) = 0,
  \label{eq_4_N}
\end{equation}
where $\Dot{N}(t)$ denotes the derivative of $N(t)$ with respect to time, $t$. As mentioned previously, here we model clogging as an instantaneous process where an open channel becomes clogged irreversibly. That is, once a channel clogs, we assume that it remains clogged for all time and no longer permits flow. Rather than working directly with $N(t)$, it is convenient to define the fraction of clogged channels $f(t)= \frac{N(t)}{N_0}$. Dividing both sides of \eqref{eq_4_N} by $N_0$ results in
\begin{equation}
\Dot{f}(t) = \lambda(q) (1-f(t)); \hspace{1cm} f(0) = 0.
    \label{mf_eq}
\end{equation}
We consider three different scenarios: devices with (i) a constant pressure difference $\Delta P$, (ii) a constant total flow rate $Q$, and (iii) independent channels with constant clogging rate $\lambda$. We obtain solutions for all these cases for single-column devices, and for multiple-column devices belonging to case (iii), by using tools from probability theory and reliability engineering \cite{reliability_eng, talbot_model, blokus2020multistate}. We expand on these results to develop a solution that tests our mean-field model for multiple-column devices for the constant flow rate devices. We first focus on the case of single-column devices in Sections \ref{subsection: p - single}--\ref{sec:const_flow - single}, then expand our results to multiple-column devices in Sections \ref{subsection: mult - constant p}--\ref{subsection: mult- const flow}. We apply the model to \cite{Du1422} subsequently in Section \ref{sec:vid_app}.

\subsection{Choosing the Clogging Rate Function}
The dependence of the clogging rate function $\lambda$ on system variables such as the flow rate through a channel $q$ is application-dependent and could take many possible forms. In the following sections, we will show the results for three representative choices of $\lambda(q)$. These functions have a simple form and remain non-negative over physical ranges of the flow rate $q$. The latter is a desired property since we limit our examples to those in which the channels cannot unclog after clogging. {To elaborate, as a first exploration of the model, we have chosen to focus on the situation in which the clogging is irreversible, but we expect that the model could be generalized to handle reversible clogging. To model reversible clogging using the same principles, it would be important to characterize the mechanism of reversibility. For instance, some experiments use oscillating flows to unclog the channels \cite{canc_sort2}. In this case, the dependence of the clogging rate {function} on the flow rate should be modified to reflect the oscillatory nature of the flow. In other examples, the unclogging could happen due to an intrinsic change in the particles as a time-dependent response to the environment conditions which could also depend on time \cite{Du1422}. In such cases, one should include a function in the formulation of the clogging rate function that captures this response of the particles.}  \par
{We assume that each cell traversing a channel has a given probability of clogging. Thus, we may write the clogging rate function in the form $\lambda = \gamma \cdot \theta$, where $\gamma$ is the rate of cells passing through a channel and $\theta$ is the probability of a cell becoming clogged. We further assume that cells are distributed uniformly throughout the fluid in the part of the device susceptible to clogging, so that $\gamma$ is proportional to the flow rate $q$ with proportionality constant given by the cell number density $\alpha$, $\gamma = \alpha q $.} \par
{The probability of clogging  $\theta$ may take different forms depending on the application. For example, previous work on deformable capsules has revealed a pass-stuck transition in the pressure-resistance phase \cite{kusters,  Bielinski_Deformbality}. This may be interpreted as a discontinuous probability function $\theta(q,r)$ that transitions from 1 to 0 at the pass-stuck boundary. Additionally, note that pressure, resistance, and flow rate across any channel are related by hydrodynamics (e.g. by the Hagen-Poisseuille equation) so that, in fact, there are only two independent parameters. For systems consisting only of parallel channels with uniform resistances, it is sufficient to only specify the channel flow rate $q$ since $p$ is a function of $q$ and the channel resistance $r$. Here, we choose $(q,r)$ to represent this two-parameter phase space.} \par 
{Decreasing the flow rate and, consequently, the shear stress in channels increases the clogging probability $\theta(q,r)$ for deformable particles to get stuck in the channels \cite{Bielinski_Deformbality, kusters}. Therefore, the overall clogging rate $\lambda = \alpha q \cdot \theta$ results from the competing effects of flow rate and clogging probability. Previous literature \cite{SauretAlban2014Cbsi} supports a linear dependence of the clogging rate on the channel flow rate for rigid particles, while for deformable particles the relationship may be more complex \cite{Lange2015_deform}. In order to illustrate the possible model behaviors, we use simple polynomial expressions in which the probability of clogging is linear or quadratic in $q$.}  \par
Although more generally, the clogging rate could depend on both the resistance of the channels $r$ and the channel flow rate as mentioned above, here we focus on its dependence on the flow rate. This follows our assumption that clogging is  an instantaneous process and that the channels are identical with fixed resistance. Note that if the key properties of the particles or cells, such as their size or deformability, were to change as a function of time or pressure, the choice of the function for the clogging rate should reflect that. For instance, the sickling of red blood cells flowing through microchannels could lead to changes in their deformability \cite{sickling_pressure}, and this should be taken into consideration when choosing the clogging rate function. Although here we focus on simple forms of $\lambda(q)$ (where we suppress the dependence on $r$ since it is assumed constant throughout the device), we emphasize that depending on the application, $\lambda$ could be adjusted and fine-tuned within the same modeling framework.     \par
As a final note on the clogging rate functions in this paper, we define them as a function of the normalized channel flow rate $\tilde{q} = \frac{q}{Q_0}$ in all our examples, where $Q_0$ denotes the initial total flow rate. In addition, we {consider a normalized} concentration of particles {(number density)} $\alpha = 1$. {Hence, the results in Section \ref{sec:results} are given in arbitrary units. To convert to physical units in an application, one would need to scale the time by a factor of $(\Bar{\alpha} Q_0)^{-1}$, where $\Bar{\alpha}$ is the number density of the cells used in that particular application.} {Appendix \ref{app:alpha} explains how we estimate the values} of $\Bar{\alpha}$ in {physical units for the experiment discussed in Section \ref{sec:vid_app}, which also serves to provide general guidelines for applying the model}. 

{While $\lambda^{-1}$ provides us with a time-scale of clogging, $(\bar{\alpha} q)^{-1}$ gives us a timescale of the arrival of particles or cells at a channel.}  {Dimensional analysis yields the dimensionless quantity $\frac{\lambda}{q \cdot \bar{\alpha}}$. This motivates choosing $ \lambda(q) = \bar{\alpha} q $ as one of the clogging rate functions explored in this paper. Additionally, since previous literature supports $ \lambda = \bar{\alpha} q $ for modeling clogging rate of rigid particles, the dimensionless quantity $\frac{\lambda}{q \cdot \bar{\alpha}}$ tells us how closely to a rigid particle the deformable particles behave.}  
\subsection{Stochastic Simulations}
We implement a time-driven stochastic model to verify our mean-field results. We initialize a matrix representing the grid structure with $n$ columns consisting of $m$ rows. Keeping the total flow or pressure difference constant, we iterate over the following procedure at each time step $\Delta t$ until all channels in one column become clogged:
\newline \begin{enumerate}
    \item Compute the flow matrix.
    \item Store the number of clogged channels $N$ and fraction of clogged channels $f$.
    \item Calculate the probability matrix from the clogging rate $\lambda (q)$ and time step $\Delta t$.
    \item Generate a random matrix of the same size and compare to values of the probability matrix. 
    \item Clog the channel if probability value for that channel is higher than the random number for that same channel. 
\end{enumerate}
{Note that this is similar to the algorithm followed in \cite{SauretAlban2018Goci} with the difference that they only consider single-column devices with the clogging rate function corresponding to $\lambda(q) = \bar{\alpha} q$. }
All the stochastic simulation results in this paper show the mean obtained from 300 trials {unless otherwise specified}. The MATLAB code to the simulations {is} available at the GitHub repository for this paper, {the link to which appears in the Data Accessibility section}. We use the standard deviation of the trials to display the error in the figures that contain stochastic simulation data sets. 
\section{Results}\label{sec:results}
\subsection{Single-Column Devices with a Constant Pressure Difference  }
\label{subsection: p - single}
\begin{figure}[h]
    \centering
    \includegraphics[width = 14cm]{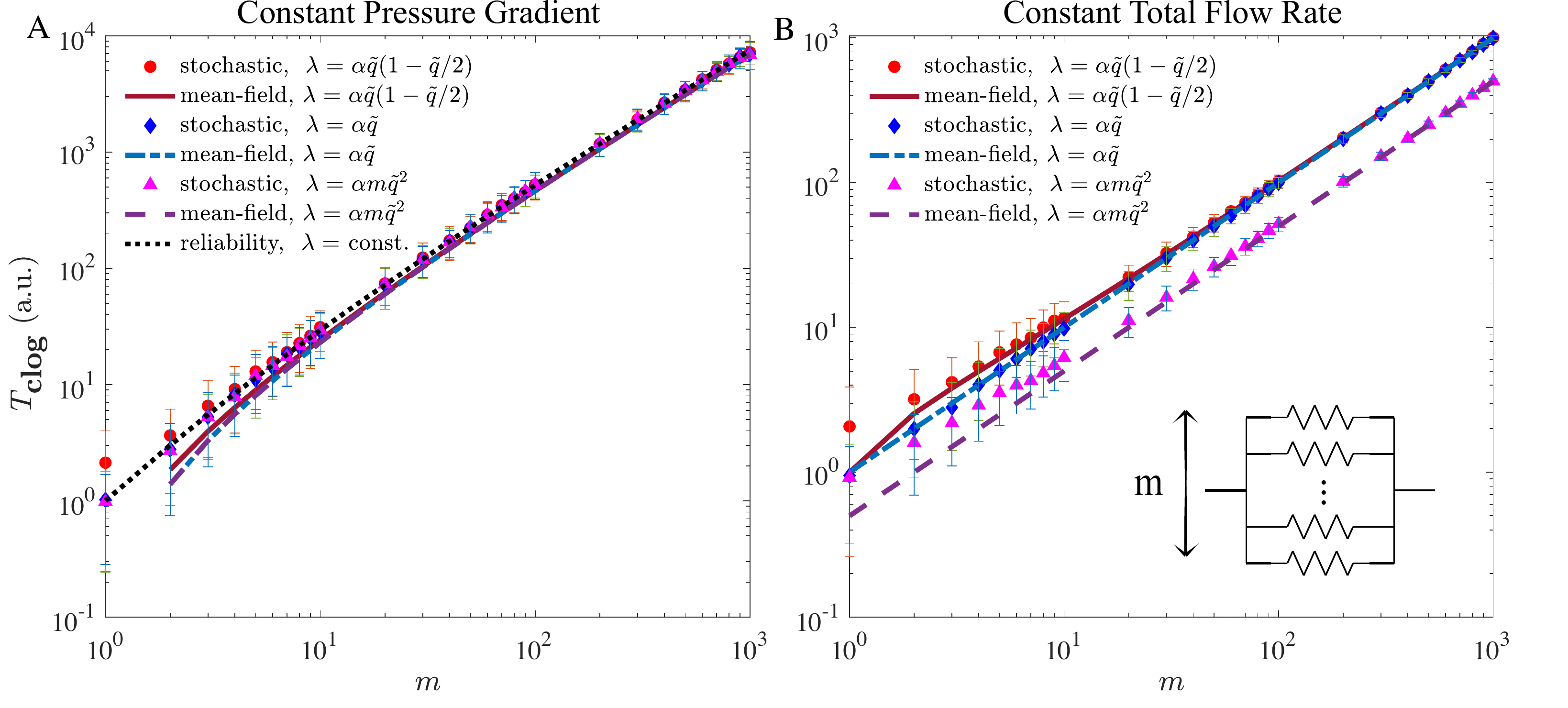}
    \caption{For a single-column device, the agreement between the mean time of clogging $T_\text{clog}$ from the stochastic simulation and the theoretical prediction from the mean-field model increases as the number of rows $m$ grows. Here $\Tilde{q} = q/Q_0$ denotes the normalized channel flow rate and $\alpha = 1$ denotes the normalized particle concentration. The error bars of the stochastic data show the standard deviation of 300 trials. \textbf{A} The single-column device with a fixed pressure offers a special case: the mean-field prediction for different choices of the clogging rate function $\lambda(\tilde{q})$ coincide when we start with the same value for $\lambda$ which remains time-independent. {Note that many of the data points coincide as the data collapses onto the same general trend.} \textbf{B} The schematic image shows a representation of the single-column device as a system of parallel resistors according to the hydraulic analogy.}
    \label{fig:cpT_m}
\end{figure}

Before considering a device consisting of multiple columns, we first apply our mean-field model to the case of devices consisting of only one column. The schematic inset in Figure \ref{fig:cpT_m} shows a resistor representation of this device. In a single-column device with a constant pressure gradient, all channels have the same flow rate. This flow rate remains constant regardless of the number of clogged channels since the pressure across each channel remains the same. Thus, a single-column device with a constant pressure difference can be modeled as having independent channels. \par
The mean-field model predicts the time of clogging given the number of channels in a row, $m$, the pressure difference $\Delta P$, and the resistance of channels $r$. Since we assume all the channels to be identical and in parallel, initially, the total resistance $R(t)$ will be a multiple of $r$, 
\begin{equation}
 R_0 \equiv R(0) = r/m,     
\end{equation}
and according to \eqref{PRQ} the initial total flow rate $Q_0$ is given by
\begin{equation}
 Q_0 = m \Delta P/r.
\end{equation}
In a single-column device, we analytically solve the mean-field equation \eqref{mf_eq}. Here, the net resistance of the system is given by $R(t) = r/(N_0-N(t))$, where $N_0 - N(t)$ is the number of open channels at time $t$ and $r$ the resistance of each channel. According \eqref{PRQ}, \begin{equation}
    Q(t) = \frac{\Delta P}{R(t)} = (N_0 - N(t)) \frac{\Delta P}{r}.
    \label{eq:(Qt1)}
\end{equation}
Since all the channels have the same resistance, the total flow is evenly distributed within all open channels. Thus, the flow through each channel $q(t)$ satisfies \begin{equation}
    q(t) = Q(t)/(N_0 - N(t)).
    \label{eq:(qt)}
\end{equation}
From \eqref{eq:(Qt1)} and \eqref{eq:(qt)} we have 
\begin{equation}
    q(t) \equiv q = \frac{\Delta P}{r}.
\end{equation}
Since we assume that the resistance of each channel and the pressure difference remain constant, $q$ and consequently the clogging rate $\lambda(q)$ do not depend on time. Therefore, with a constant $\lambda$, we expect the results to be identical to the independent channels case. \par
In this case, we may analytically solve the mean-field \eqref{mf_eq} using integration by parts to get
\begin{equation}
\label{eq:asymp}
    f(t) = 1 - e^{-\lambda t}.
\end{equation}
This allows us to predict the time of clogging $T_\text{clog}$, {defined as the time at which all channels in at least one column have clogged}. {Since we are considering a single-column device here, all channels in the device clog when $t = T_\text{clog}$ implying} $f(T_\text{clog}) = 1$. Since \eqref{eq:asymp} only reaches 1 asymptotically, we approximate the clogging time as the instance at which only 1 open channel remains, \begin{equation}
   f({T_\text{clog}}) \geq \frac{m-1}{m}, 
\end{equation}
to arrive at 
\begin{equation}
    T_{\text{clog}} \geq \frac{\ln{(m)}}{\lambda},
\end{equation}
where we have used \eqref{eq:asymp}.
This approximation agrees with the theoretical results $T_\text{clog}$ {from reliability methods (Appendix \ref{Ap:Ind_channels})} up to the first order \cite{talbot_model}:
\begin{equation}
    T_\text{clog} = \frac{1}{\lambda} H_m,
\end{equation}
where $H_m$ is the $m$th Harmonic number. At the limit where $m$ approaches infinity,
$$ H_m \xrightarrow{} \ln{m} + \gamma + O(\frac{1}{m}),$$
with $\gamma$ the Euler-Mascheroni constant. \par
Figure \ref{fig:cpT_m}\textbf{A} shows the results comparing the mean-field prediction of clogging time $T_\text{clog}$ as a function of $m$ for three different clogging rate functions $\lambda(\tilde{q})$:  $\lambda(\tilde{q}) = \alpha \tilde{q} $,
     $\lambda(\tilde{q}) = \alpha m  \tilde{q}^2 $,
    and $\lambda(\tilde{q}) = \alpha \tilde{q} (1 - \tilde{q}/2) $. 
Here, $\tilde{q} = \frac{q}{Q_0}$ is the normalized channel flow rate, and $\alpha = 1$ denotes the normalized particle concentration. To allow for a fair comparison between these cases, we set the initial value of the clogging rate to be approximately equal. This sets the value of the multiplicative prefactor in the preceding expressions for $\lambda(\tilde{q})$. {As expected, the mean-field model breaks down in the limit of small number of rows. In the limit $m = 1 $ corresponding to a single channel, the clogging time is not defined in the mean-field model. For this reason, we do not plot this in the figure.}  \par
In addition, this figure contains the results from the stochastic simulation for each of the mean-field curves and analytic predictions from reliability engineering (see Appendix \ref{Ap:Ind_channels} for the calculations). As the figure shows, the agreement the stochastic simulation, {reliability} theory, and the mean-field model increases with the number of channels. \par
\subsection{Single-Column Devices with a Constant Total Flow Rate}
\label{sec:const_flow - single}
Here, we provide the parallel analysis to Section \ref{subsection: p - single} for the case of devices with a constant total flow rate. Given a constant total flow rate $Q_0$ through each column and $m$ channels in a column, we model the fraction of clogged channels $f(t)$ using \eqref{mf_eq}, the mean-field equation. The inset in Figure \ref{fig:cpT_m}\textbf{B} shows a resistor circuit representation of the single-column device with $m$ rows. We analytically solve \eqref{mf_eq} {for the clogging time} with the same clogging rate functions used previously. \par
First, we consider a class of clogging rate functions where $\lambda(\Tilde{q}) = \alpha \tilde{q}^\beta  $ and $\beta$ is a non-negative integer, with $\alpha $ representing the normalized number density of the particles, and $\tilde{q} = \frac{q}{Q_0} $ denoting the normalized flow rate through each channel. Note that, unlike the case of constant pressure difference, in this case the channel flow rate $q=q(t)$ varies with time even in the case of a single column. For constant $\lambda$,  ($\beta = 0$), the solution to \eqref{mf_eq} is given by $f(t) = - e^{-\lambda t} + 1$, similar to the results discussed for a single-column device with a constant pressure difference {in Section \ref{subsection: p - single}}. When $\beta \geq 1$,  we can write the flow through each channel as
\begin{equation}
\label{lilq}
    q(t) = \frac{Q_0}{m-N(t)}, 
\end{equation}
where $N$ denotes the number of clogged channels.
Dividing both sides of \eqref{lilq} by $Q_0$, we have
\begin{equation}
\label{eq:Qq}
    {\tilde{q}(t)}{} = \frac{1}{m-N(t)} = \frac{1}{m(1-f(t))}.
\end{equation}
We now may write \eqref{mf_eq} as    
\begin{equation}
\label{diff1}
    \dot{f}(t) = \alpha \tilde{q}(t)^\beta  ({1-f(t)}) = \alpha' (\frac{1}{1-f(t)})^\beta (1-f(t)),
\end{equation}
where we have used \eqref{eq:Qq} to arrive at the final expression  and $\alpha'  \equiv \frac{ \alpha}{m} $. Using integration by parts, for a single-column device, we have
\begin{equation}
\label{eq:sol_beta}
   - \frac{(1 - f(t))^\beta}{\beta} + \frac{1}{\beta} = \alpha' t.
\end{equation}
Now, we calculate the clogging time by evaluating \eqref{eq:sol_beta} when $f(T_\text{clog}) = 1$:
\begin{equation}
\label{betasol}
     T_\text{clog} = \frac{1}{\beta \alpha'}.
\end{equation}

To illustrate further with another example, we evaluate $f(t)$ for another choice of the clogging rate function, 
\begin{equation}
\label{testlamq1-q}
    \lambda(\Tilde{q}) = \alpha \Tilde{q} (1-\frac{\Tilde{q}}{2}).
\end{equation}
Rewriting \eqref{testlamq1-q} using \eqref{eq:Qq} and then substituting it in \eqref{mf_eq} results in
\begin{equation}
    \dot{f}(t) = \alpha'   (1-\frac{1}{2m(1-f(t))}).
\end{equation}
For convenience, we define the small parameter $\epsilon \equiv \frac{1}{2 m}$ to write
\begin{equation}
\label{diff_i}
    \dot{f}(t) = \alpha' (1-\frac{\epsilon}{1-f(t)}).
\end{equation}
For a single-column device, solving \eqref{diff_i} by separation of variables, we arrive at a Lambert W function, ${{\textrm{W}}}_0$:
\begin{equation}
   f(t) =  1-\epsilon \,{{\textrm{W}}}_0 \left(\frac{{ \,{1-\epsilon }} }{\epsilon } e^{\frac{1 - \epsilon}{\epsilon }- 2 \alpha' t }\right)-\epsilon,
\end{equation}
where $f(t)$ satisfies
\begin{equation}
\label{sol_f_Q}
    f(t) - \epsilon \ln\left|1- f(t) -\epsilon\right|  = \alpha' t + \epsilon \ln\left(1-\epsilon\right).
\end{equation}
When the device {only has one column, it} fails when $f(T_{\text{clog}}) = 1$. So by \eqref{sol_f_Q}:
\begin{equation}
  T_{\text{clog}} =  \frac{m}{ \alpha} - \frac{1}{2 \alpha}  \ln\left( \epsilon (1-\epsilon)\right).
\end{equation}

Figure \ref{fig:cpT_m}\textbf{B} shows the clogging time results for this choice of clogging rate function \eqref{testlamq1-q} and for the class of clogging rate functions discussed above when we have one column, $n =1$. The mean-field model agrees with the stochastic simulations when predicting the mean time of clogging. Similar to the cases with a constant pressure difference, the agreement increases when we have a larger number of rows. \par
The case of the constant total flow rate appears in a wide variety of studies concerned with the reliability of systems. These include studies of constant flux in bundles of channels and understanding the symmetric load sharing in materials \cite{fbm1} in addition to resistor circuits. Thus, as we will show, we may expand upon previous results \cite{talbot_model} in order to obtain benchmark solutions for our problem in certain special cases. For a single-column device, \cite{talbot_model} offers a formalism to find the clogging time and the fraction of clogged channels. This allows us to confirm our results from the mean-field model. \par

\subsection{Multiple-Column Devices with a Constant Pressure Difference}
\label{subsection: mult - constant p}
Consider a device with a fixed pressure difference $\Delta P$ and $n$ columns. In this section, we generalize the results from Section \ref{subsection: p - single} in which $n=1$. We repeat our analysis in the case of $n>1$ by solving the corresponding mean-field equations and comparing to the results of the stochastic simulations. Initially, each column has $m$ open channels, each with a resistance $r$. 
\begin{figure}[h]
    \centering
    \includegraphics[width = 11cm]{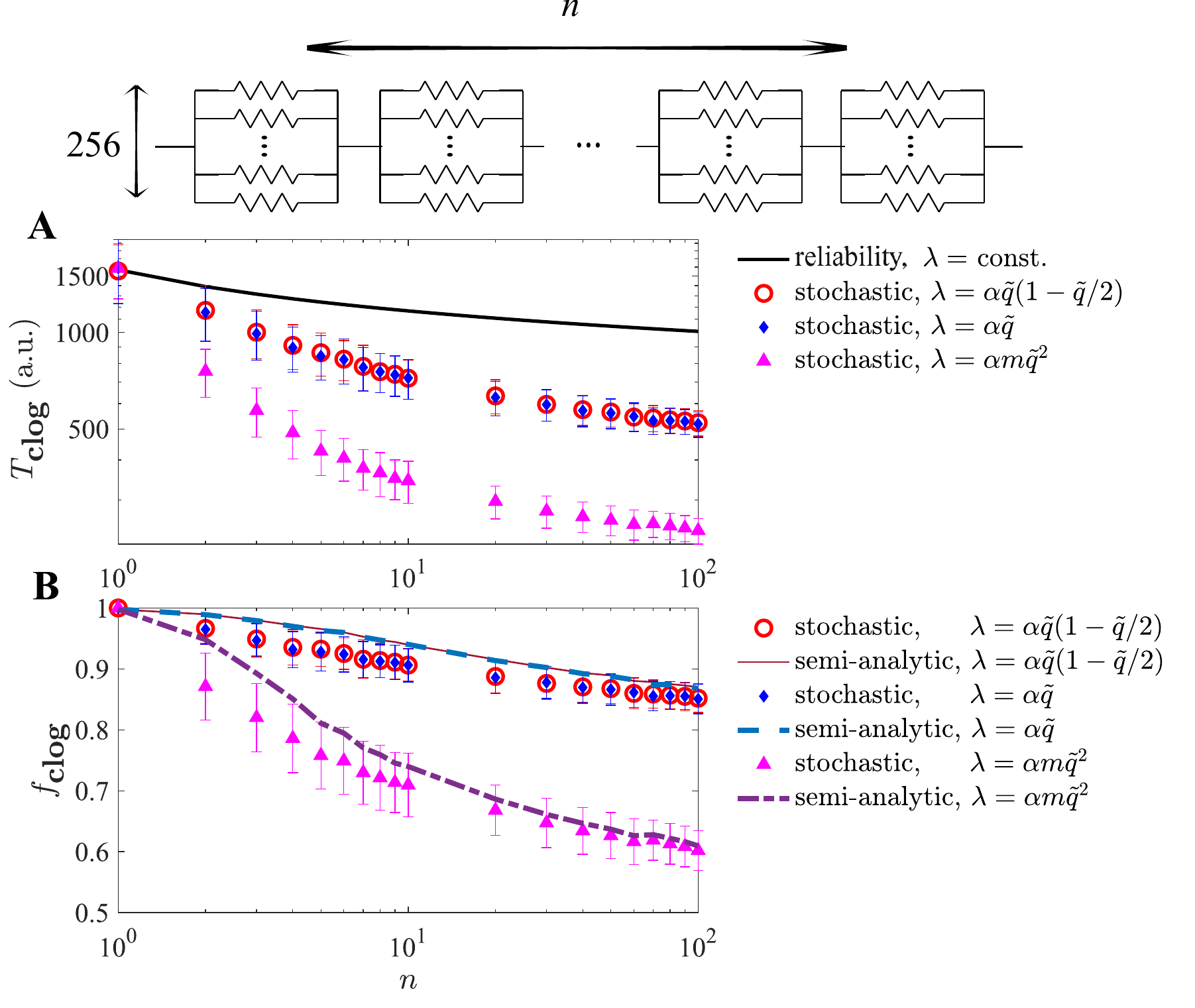}
    \caption{The mean-field model helps predict the mean time of clogging indirectly for a device with multiple columns and a fixed pressure. \textbf{A} For the choices of $\lambda(\tilde{q})$ presented in the figure, the mean-field prediction provides an upper bound on the mean time of clogging as a function of number of columns $n$ for a device with a constant pressure difference. \textbf{B} The fraction of clogged channels at the time of clogging from the stochastic simulations agrees with the semi-analytical data which is obtained by using the mean clogging time from the stochastic simulations in the mean-field equation.  The schematic image shows a representation of a device with $n$ columns and $m = 256$ rows as a resistor system. The error bars display the standard deviation of 300 trials.}
    \label{fig:cpf_T}
\end{figure}
The schematic image in Figure \ref{fig:cpf_T} illustrates a resistor representation of a multiple-column device with $n $ columns and $m$ rows. 
As above, we assume that the clogging rate function of the $i$th channel $\lambda_i$ depends on the flow through that channel $ q_i$. \par
First, $R(t)$ the total resistance of the device at time $t$ depends on the number of open channels in each column as follows:
\begin{equation}
\label{eq: R - multiple p}
    R(t) = \sum_{i=1}^n R_i (t) = \sum_{i=1}^n \frac{r}{  {m - N_i(t)}}  = \frac{r}{m} \sum_{i=1}^n \frac{1}{  {1 - f_i(t)}},
\end{equation}
where $N_i(t)$ and $R_i(t) $ respectively denote the number of clogged channels and net resistance in column $i$ at time $t$. Here,
\begin{equation}
    f_i (t) = \frac{N_i(t)}{m}
    \label{eq:def_fi}
\end{equation}
denotes the fraction of clogged channels in column $i$, and $f(t) = \frac{N (t)}{n m}$ denotes the fraction of clogged channels in the entire device.  
    Using equations \eqref{eq: R - multiple p} and \eqref{PRQ}, the total flow in the device $Q (t)$ and the total resistance $R (t)$ are related to the total pressure difference $\Delta P$ by
\begin{equation}
\label{eq_Q}
    Q (t) = \frac{ \Delta P}{(r/m) \sum_{i=1}^n  \frac{1}{1 - {{f_i} (t)}}}.
\end{equation}
By the conservation of mass, the flow in each row of the $j$th column satisfies 
\begin{equation}
\label{eq:f_jq_j}
    q_j (t) = \frac{Q (t)}{m - N_j (t)} = \frac{Q (t)}{m (1-f_j (t))},
\end{equation}
where we have used \eqref{eq:def_fi}.
Rewriting \eqref{eq:f_jq_j} using \eqref{eq_Q}, we have 
\begin{equation}
    q_j (t)  = \frac{\Delta P}{r\sum_{i=1}^n  \frac{ 1-f_j(t) }{ 1-f_i(t)}}.
     \label{eq:q}
\end{equation}
Given that initially all the channels are open, we have $f_j (0) = f (0) = 0$ and
\begin{equation}
  q_j (0) = \frac{\Delta P}{r} \left[   \frac{1}{n} \right] ; \hspace{1cm} Q (0) = \frac{\Delta P}{r} \left[   \frac{m}{n} \right].  
\end{equation}
We may write \eqref{mf_eq}, the mean-field equation,  for column $j$ as
\begin{equation}
\label{eq:mf-multiplecolumns}
    \Dot{f_j }(t) = \lambda(q_j) (1- f_j(t)).
\end{equation}
Since all of the columns have the same number of rows in the grid-like geometry considered, $f(t) = \sum_{i=1}^n \frac{f_i(t) }{n}$, and
\begin{equation}
    \Dot{f} (t) = \sum_{i=1}^n \frac{\Dot{f_i} (t) }{n}.
\end{equation}
Because initially all channels are open so that $f_i(0) = 0$ for all $i$, $\lambda(q_i(0)) \equiv \lambda(\frac{\Delta P}{n r}) $, and without any stochastic effects, we expect all {columns} to proceed identically and clog at the same time due to the symmetry of this problem. Thus, the clogging time{, $T_\text{clog}$ the time at which all channels in at least one column have clogged,} would be the same as in the case of the single-column device. In a real device, however, the stochastic effects break this symmetry and affect the clogging time.

{The stochastic jumps in Figures \ref{fig:data_col_cp} and \ref{fig:dat_col_cf} highlight why a mean-field model fails to capture the time of clogging in the case of multiple-column devices. The deterministic model treats all columns equally, whereas stochastic effects create an asymmetry in how columns progress in their clogging over time. The stability analysis, further expanded on in Appendix \ref{Ap:stability_P} and \ref{Ap:stability_Q}, helps us understand how significant the effect of this asymmetry is and how much sooner the first column clogs compared to the case in which all columns clog at the same time.}

{Although the definition for $T_\text{clog}$ remains unchanged in the passage from single-column to multiple-column devices, the actual clogging dynamics can be very different. In both cases, by using the mean-field equation, one can solve for $T_\text{clog}$ when the fraction of clogged channels at the time of failure $f_\text{clog} = f(T_\text{clog})$ is known. In contrast to the case of single-column devices where $f_\text{clog}$ is trivially known and equal to 1, in multiple-column devices, $f_\text{clog}$ can attain a range of values governed by the stochastic effects as shown in Figures \ref{fig:cpf_T}\textbf{B} and \ref{fig:cf_Tf}\textbf{B}.} Since the {solution to the} mean-field equation does not capture the stochastic effects in multiple-column devices, it only provides limited information on the time of clogging {when $f_\text{clog}$ is unknown}. \par 
Figures \ref{fig:cpf_T}\textbf{A} and \ref{fig:cpf_T}\textbf{B} show the mean time of clogging $T_\text{clog}$ and the fraction of clogged channels at the time of clogging $f_\text{clog} $ as a function of number of columns $n$, respectively. {The error bars in these figures show standard deviation of the 300 stochastic simulation trials used for the results shown in the figure.} As evident from Figure \ref{fig:cpf_T}\textbf{B}, the fraction of clogged channels at the time of clogging  does not attain $f = 1$ for devices with two or more columns. This is because the device fails as soon as the \emph{first column} completely clogs, which typically occurs before all of the channels in the device clog. Thus, the time at which the first column fails determines the time of device failure. \par
Although {solving} the mean-field equation does not directly lead to the time of failure in the multiple-column devices with a constant pressure gradient, it may be combined with other methods to yield an estimate of the failure time. For instance, to obtain the semi-analytical values of $f_\text{clog}$ shown in Figure \ref{fig:cpf_T}\textbf{B}, we first compute the mean time of clogging, $T_\text{clog}$ using stochastic simulations and then insert this value into the solution to the mean-field equation \eqref{mf_eq}. We can similarly extract the corresponding mean time of clogging using the mean-field equation to find the time at which the solution curve attains the value $f_\text{clog}$ for instances when this information is available by other means. {Another way to assess the stochastic effects on $T_\text{clog}$ when going from single-column to multiple-column devices is by stability analysis. {See Appendix \ref{Ap:stability_P} and \ref{Ap:stability_Q} for further discussion.}} \par
\subsection{Multiple-Column Devices with a Constant Total Flow Rate}
\label{subsection: mult- const flow}
\begin{figure}
    \centering
    \includegraphics[width = 12cm]{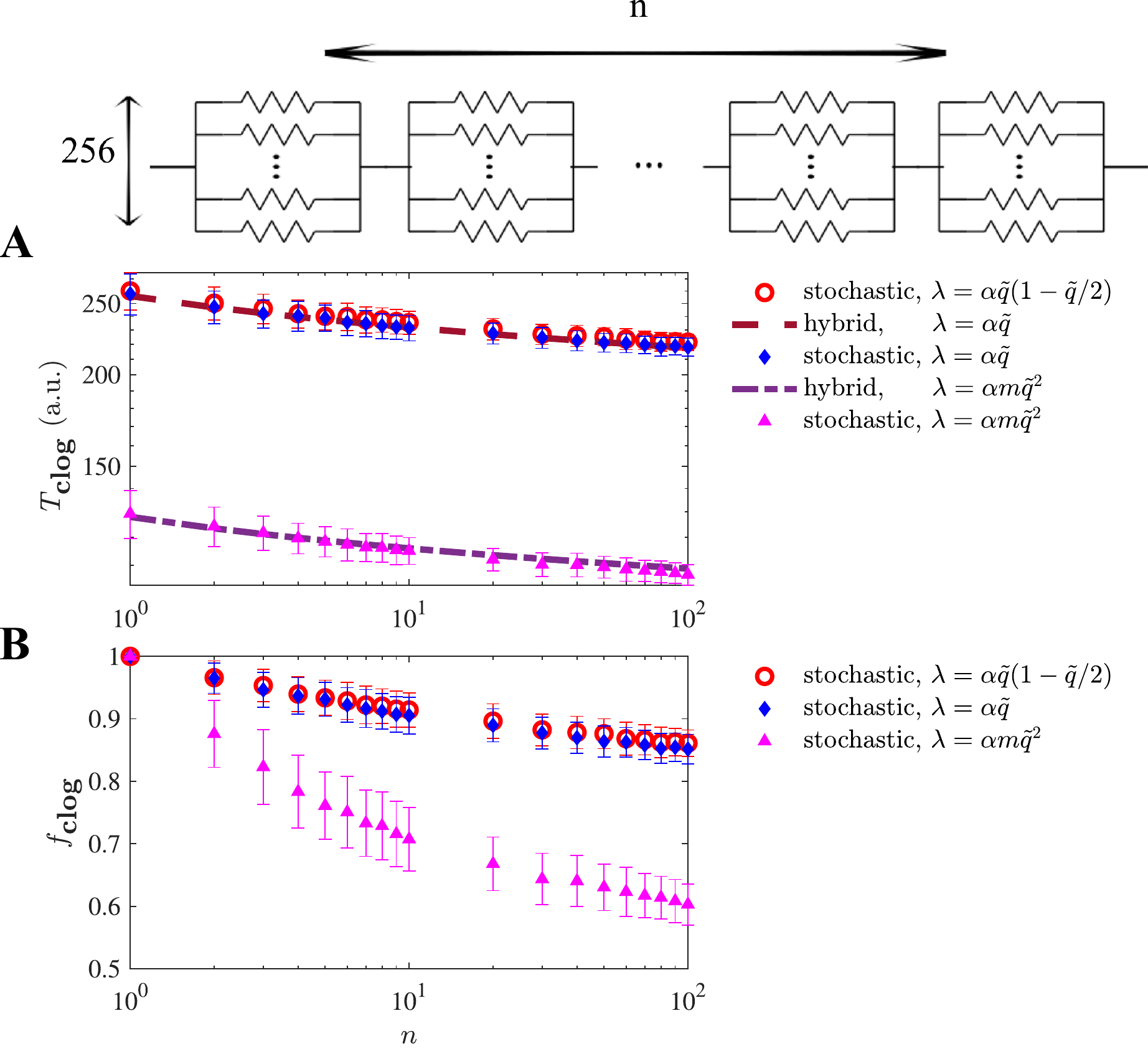}
    \caption{In multiple-column devices, the device clogs before all the channels clog, and we need other tools in addition to the mean-field model to predict the time of clogging. \textbf{A}  mean time of clogging   \textbf{B} and the fraction of clogged channels at the time of clogging as a function of number of columns $n$ for a device with a constant total flow rate, $Q_0$. The different curves correspond to different clogging rate functions, $\lambda(\tilde{q})$, as a function of the normalized flow $\Tilde{q} = \frac{q}{Q_0}$ and normalized particle concentration $\alpha = 1$. The schematic image shows a resistor representation of a device with $n$ columns and $m = 256$.  
All stochastic data points show the mean collected over 300 trials, and the error bars display the standard deviation. The hybrid data points result from solving \eqref{theorycf} numerically using Mathematica for the clogging rate functions $\lambda = \alpha \Tilde{q} $ and $\lambda = \alpha \Tilde{q}^2 $.}
    \label{fig:cf_Tf}
\end{figure}

For the case of multiple-column devices with constant total flow rate, we may calculate the mean time to clogging $T_\text{clog}$ by using the formalism in \cite{talbot_model} combined with the observation that the columns are independent. Figure \ref{fig:cf_Tf}\textbf{A} shows a resistor representation of such a device with $n$ columns and $m = 256$ rows. The mean time of clogging of the first column out of the $n$ columns satisfies
\begin{equation}
\label{tpsmq}
    T_\text{clog} = \int_0^\infty ({R_p}(t))^n dt,
\end{equation}
where {$R_p$} is the {reliability} function for a set of $m$ parallel channels {(see Appendix \ref{Ap:Ind_channels})}. When $\lambda(\tilde{q}) = \alpha \tilde{q}$, by \cite{talbot_model} and \cite{blokus2020multistate}, we have 
\begin{equation}
\label{survivalf}
    {R_p}(t) = e^{-m \lambda_0 t }\sum_{k = 0}^{m-1} \frac{(m \lambda_0 t )^k}{k!},
\end{equation}
where $\lambda_0${=$\alpha/m$} denotes the clogging rate function at time $t = 0$ when $\tilde{q}(0) = \frac{1}{m}$.  Eq.~\eqref{survivalf} can be written in terms of Gamma and incomplete Gamma functions \cite{JamesonGIncGamma}:
\begin{equation}
\label{psmq}
    {R_p}(t)  = \frac{\Gamma(m, m \lambda_0 t )}{\Gamma(m)},
\end{equation}
{where $R_p(t)$ may be recognized as the regularized (or normalized) Gamma function.} Putting \eqref{psmq} and \eqref{tpsmq} together, we have
\begin{equation}
 T_\text{clog} = \int_0^\infty \left(\frac{\Gamma(m, m \lambda_0 t )}{\Gamma(m)}\right)^n dt.
    \label{theorycf}
\end{equation}
The inset in Figure \ref{fig:cf_Tf}\textbf{A} shows the mean time of clogging from the stochastic simulation compared with the theoretical prediction calculated using \eqref{theorycf}. The formula \eqref{theorycf} yields a solution in terms of an integral, which may be solved analytically for $n=1,2$ and which we solve numerically for larger values of $n$. {We note that since the regularized Gamma function is non-negative and bounded from above by 1 for the physical values of $t$ and $m$ considered here, increasing $n$ leads to a smaller $T_\text{clog}$. This is also evident in simulations as seen in Figure \ref{fig:cf_Tf}\textbf{A}.}
To obtain the theoretical results shown in Figure \ref{fig:cf_Tf}\textbf{A} for $\lambda(\tilde{q}) = \alpha m \tilde{q}^2 $, we use the observed time of clogging for each column \eqref{betasol} along with \eqref{theorycf}. To elaborate, when $\beta = 1$, we calculate the clogging time for each column individually $T_{\text{clog}, \beta = 1} =\frac{1}{\alpha'} $ from \eqref{betasol}. This clogging time scales linearly with the clogging time when $\beta = 2$, i.e.
$$T_{\text{clog}, \beta = 2} =\frac{1}{2 \alpha'} T_{\text{clog}, \beta = 1} = \frac{1}{2 \alpha \lambda_0}. $$
Using this observation, we modify the {reliability} function \eqref{psmq} in \eqref{theorycf}
\begin{equation}
    P_{Sm, \beta= 2}(t)  = \frac{\Gamma(m, m (2 \alpha \lambda_0) t )}{\Gamma(m)}.
\end{equation}
to obtain the results for $\lambda(\tilde{q}) = \alpha m \tilde{q}^2 $.  \par

\subsection{Comparison Between the Constant Pressure Gradient and Constant Total Flow Devices}
{In devices with a single column $n = 1$ and $m$ rows,} similar to the case of constant pressure gradient, the mean-field equation \eqref{mf_eq} predicts the time of clogging for a device with a constant total flow rate. Figures \ref{fig:cf_Tf} and \ref{fig:dat_col_cf} show our results analogous to Figures \ref{fig:cpf_T} and \ref{fig:data_col_cp}  respectively. In contrast to the case of constant pressure difference, the flow through each channel depends on the number of open channels when we have a constant total flow. Consequently, we see a more variety of functions as solutions to \eqref{mf_eq} depending on the choice of $\lambda(q)$ compared to only one form of solution \eqref{eq:asymp} for the case of constant pressure difference.   \par

{The multiple-column devices manifest a broader range of dynamics due to the competition in clogging between the columns. Since the device failure time is determined by the minimum failure time among all the columns, the failure time of the device depends on extreme-value statistics rather than depending on the mean. Thus, although the mean-field model accurately predicts the progression of the mean fraction of clogged channels up to the failure time of the device, one would need additional information to accurately predict the time of clogging. In both cases of constant pressure gradient and total flow, the solution to the single-column case $n = 1$ provides an upper bound on the mean time of failure of device $T_\text{clog}.$ Again, this arises due to the dependence of the multiple-column device failure time on the minimum value and not the mean of the column failure times. Thus, predicting the multiple-column device failure time requires access to higher order statistics of the problem and not just the mean. We note that as we have demonstrated (e.g., see Section \ref{subsection: mult- const flow}), the methods in \cite{talbot_model} for solving for the failure time may be expanded for applications to multiple-column devices.} 

{D}epending on the choice of the clogging rate function, we see different types of feedback for the two cases of constant total flow rate and constant pressure difference. For instance, when $\lambda(\tilde{q}) = \alpha \tilde{q}^2,$ we observe a {positive feedback and when $\lambda(\tilde{q}) = \alpha \tilde{q}$ a neutral feedback  for both.} As $n$ increases, $T_\text{clog}$ decreases, and the type of feedback determines how dramatic this decline is. In devices with positive feedback, we expect a larger decline in $T_\text{clog}$ and $f_\text{clog}$ due to stochastic jumps as $n$ grows, a smaller decline in devices with negative feedback, and somewhere in between the two in devices with neutral feedback. {See Appendix \ref{Ap:stability_P} and \ref{Ap:stability_Q} for further discussion on stability analysis and examples of different types of feedback.}  

\subsection{Application of the Model}
\label{sec:vid_app}
We next apply our mean-field model to experimental data from \cite{Du1422} in which sickled red blood cells cause a microfluidic device to clog over time. {Although we have presented the results in arbitrary units so far, in this section we use the physical parameters of the experiment to obtain results in SI units}. {Although the complete video consists of both clogging and unclogging phases, we only model the clogging phase here and use the corresponding portion of the video from time $t = 21.0s$ to $t = 37.5 s$. This justifies our use of a strictly positive clogging rate function corresponding to irreversible clogging over the time interval of interest in the experiment.}
\begin{figure}
    \centering
    \includegraphics[width = 12cm]{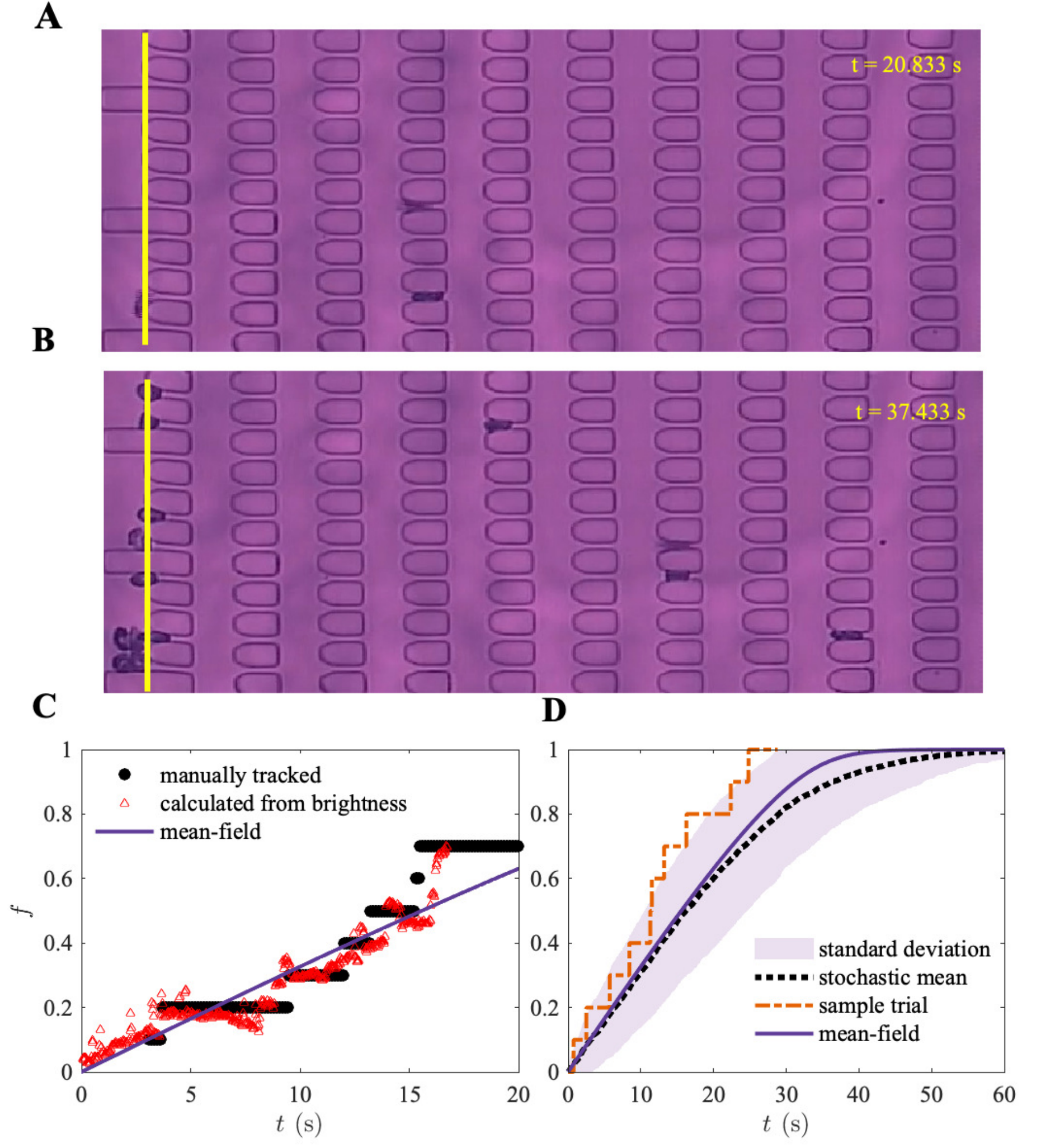}
    \caption{We apply the mean-field model to experimental data from \cite{Du1422} involving the clogging of a cell sorting device over time. \textbf{A} {A snapshot of the video at time $t = 20.833 s$.} We use a line profile (illustrated in yellow) to track the brightness of each video frame at the entrance of the clogging column from time $t = 21.0s$ to $t = 37.5 s$.\textbf{B} {A snapshot of the video at a later time $t = 37.433 s$ shows that most of the clogging occurs at the first column.} \textbf{C} {The fraction of clogged channels $f$ extracted from the brightness plot agrees with the mean-field model. The clogging phase of the video ends before $f$ reaches $1$ as also seen here.} \textbf{D} Mean fraction of clogged channels at the time of clogging computed from 1,000 trials using the clogging rate $\lambda(q) =  c_b \alpha q $ with the coefficient $c_b$ obtained from fitting. The dash-dotted line plots a single realization of the stochastic simulation, which exhibits a step-like behavior reminiscent of the experimental data.
}
\label{fig:vid_app}
\end{figure}
As Figure \ref{fig:vid_app}\textbf{A} shows, we use a line profile (in MATLAB) to track the brightness at the entrance of the first column, which we will refer to as column 1, as this column starts to clog. Since almost all the clogging occurs in this column for the duration of our analysis, we use our mean-field equation to model the fraction $f(t)$ of clogged rows in this column with respect to time. {However, this does not mean that we can treat the device as a single-column device. Despite the fact that clogging occurs nearly exclusively in a single column, it is important to model the multiple-column geometry. To correctly capture the progress of clogging here, one must consider the resistance of the device as a whole.} As the channels in this column clog, the total resistance $R(t)$ increases such that
\begin{equation}
\label{eq:vidapp_R}
    R(t) = 9 \frac{r}{10} + \frac{r}{10(1-f(t))},
\end{equation}
where we have $n = 10$ columns in total, each with identical resistance $r$. Using \eqref{PRQ} and \eqref{eq:vidapp_R}, we calculate the total flow rate and the channel flow rate $q$. Assuming the following simple form of the clogging rate function: 
\begin{equation}
\label{eq:vidapp_lam}
    \lambda(q) = c_b \alpha q,
\end{equation}  
where $\alpha$ is the cell concentration and $c_b$ is a {dimensionless} scaling coefficient obtained from the brightness data, we can solve the mean-field equation \eqref{mf_eq} for column 1. By separation of variables, we arrive at the implicit solution for $f(t)$:
\begin{equation}
\label{eq:vidapp_impsol}
 t = \frac{r}{c_b \Delta P \alpha } \left( 9 f(t) - \log(1-f(t)) \right).   
\end{equation}
{To calculate $f(t)$, we need the values for the parameters that appear in $\eqref{eq:vidapp_impsol}$, $r, \Delta P, \alpha,$ and $c_b$. Table \ref{tab:variables} gives the values for $\Delta P$ and $r$ in SI units according to the information available from the experiment. To estimate $\alpha$, we use the known concentration of cells in blood of $5 \times 10^{15} \text{cells}/m^{3}$ \cite{cell_conc}. To calculate the fitting parameter $c_b$, we can either track cells manually or use the brightness of the video frames.} As the cells increasingly clog more channels, the total brightness decreases and we may interpret the clogging events by dips in the total brightness. We assume the decrease in percentage of brightness to be proportional to the increase in the fraction of clogged channels. {We estimate the coefficient $c_b = 0.001701$ using a fitting function in MATLAB (see Data Accessibility section for accessing the code) to fit \eqref{eq:vidapp_impsol} to the brightness data}.   \par
Next, we set up the stochastic simulation using this clogging rate function. Figure \ref{fig:vid_app}\textbf{D} displays the results from both the stochastic simulation and the mean-field model. As Figures \ref{fig:vid_app}\textbf{C} and \textbf{D} shows, the mean-field model successfully captures the progression of the fraction of clogged channels with respect to time. This provides a proof of principle that our mean-field model may be used to predict the clogging progression over time in a real-world device. \par

\section{Discussion}
Clogging in microfluidic devices has important consequences for their behavior and performance in a variety of applications. Our mean-field model provides a minimal and simple way of understanding this problem for a class of cell sorting devices given significantly few quantities including the pressure difference or flow rate, general geometry of the channels, and  fluid properties. \par
We use a time-driven stochastic simulation to verify our mean-field model for a single-column device and explore its limitations in a multiple-column device. Using our model, we successfully predict the time of clogging for the single-column devices in all of the three cases that we consider: constant pressure gradient, constant total flow rate, and independent channels. For the cases of independent channels and constant flow rate, we also confirm the time of clogging using theoretical predictions from reliability engineering and probability theory. For devices with multiple columns, our model allows us to indirectly calculate the time of clogging given the fraction of clogged channels at the time of clogging. \par
As a proof of principle, we have applied the mean-field model to experimental data of clogging in a cell sorting device. Extracting the relevant clogging function leads to a prediction of how the material properties of cells and their rate of clogging change in response to sickling. This suggests that our model may be able to predict the clogging dynamics of microfluidic devices in other contexts as well. \par
Although our model predicts the behavior of ideal simple devices with instantaneous clogging, it fails to capture the spatial correlations and particle-particle and particle-pore interactions. By hydraulic analogy, we assume negligible cooperativity among multiple cells. An interesting future direction would be to explore how nearest-neighbor interactions between the channels could affect the clogging time. In some cases, we could expect an avalanche of failures and, consequently, a decrease in the clogging time when a limited number of neighbors of a channel share the extra flux due to its clogging, similar to the propagation of fractures in materials with local load-sharing \cite{perc_materials}. \par
In this paper, we have focused on the case in which all particles approximately have the same physical properties, including size, with small fluctuations due to heterogeneity. In more general filtration applications, however, the particles may vary in size, some being smaller than the channels. In addition, we only focus on scenarios where the ratio of pore to particle size is smaller than one and do not consider collective clogging due to aggregation or bridging where multiple particles block a channel or pore \cite{softmatter}. The model could be extended to consider these more complex scenarios.  \par
\begin{figure}
    \centering
    \includegraphics[width = 12cm]{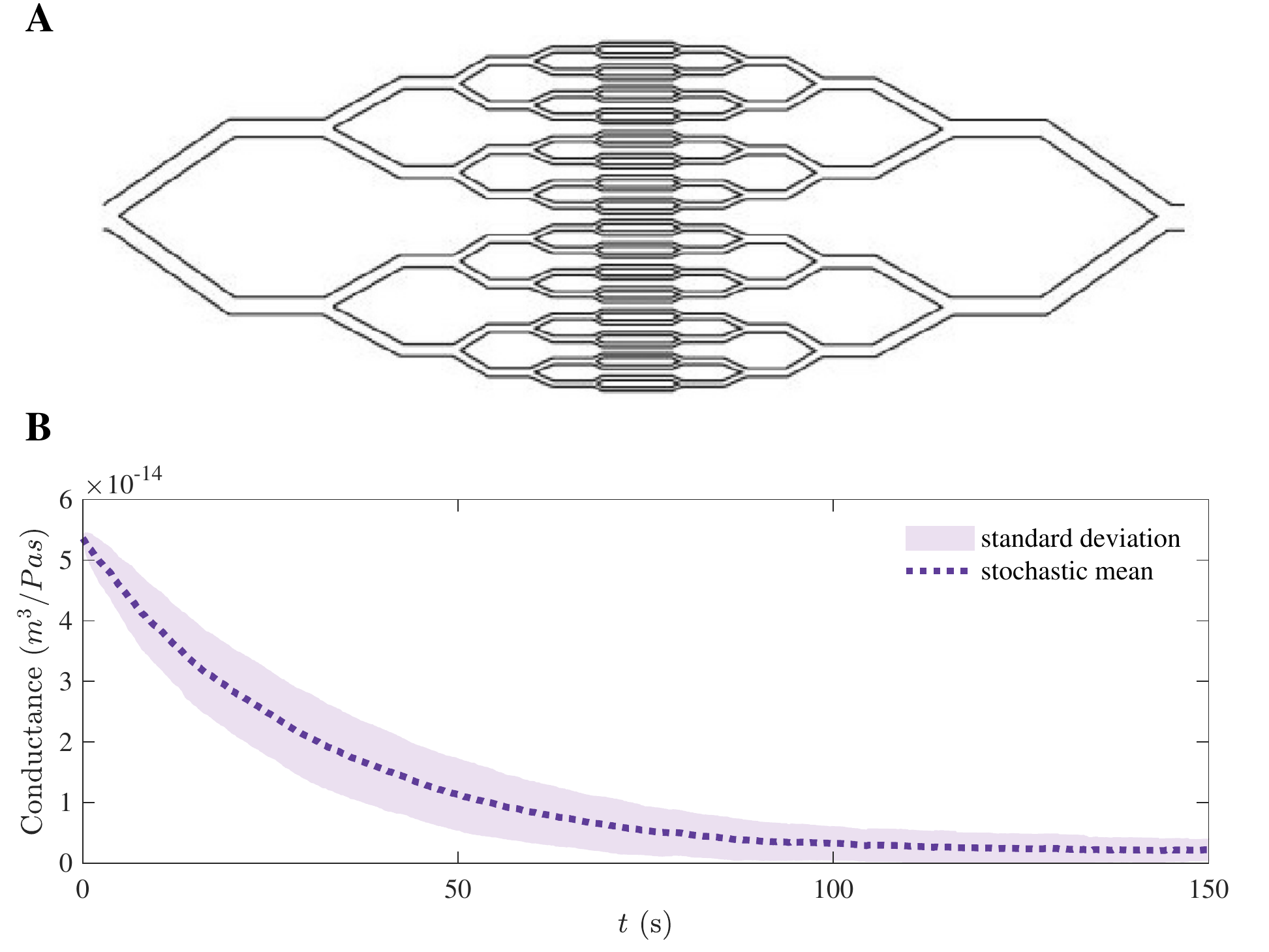}
    \caption{{We can use stochastic simulations to predict clogging behavior in more complex networks following the same modeling principles introduced in this paper. \textbf{A} We simulated the narrowest channels in this design taken from \cite{TsaiMichelle2012Ivmo} as $4\mu m$ wide to ensure sieving. \textbf{B} The conductance (inverse resistance) calculated from simulating clogging in this network decreases as the channels clog.}}
    \label{fig:complex_netw}
\end{figure}
Another potential application of our model involves more complex geometries. {As an illustrative example, in Figure \ref{fig:complex_netw}, we have shown an application to a model capillary network. This layout differs significantly from a grid as it includes channel bifurcations. The same modeling principles generalize and we may numerically investigate the properties of these more complex systems. However, formulating the mean-field equation may not be as straightforward in this case.} To apply the model to a wider range of networks, it may be helpful to identify the network parameters that influence the clogging rate function significantly and further understand their role in clogging. This way, instead of directly relying on the geometry of large and complex networks, a few parameters, which are not limited to but may include the smallest eigenvalue of the normalized network Laplacian or the diameter of the network, would sufficiently characterize the clogging. \par
Lastly, in this paper, we focus on examples where we explore the dependence of the clogging rate function $\lambda(q,r)$ on the channel flow rate $q$. We note that for more complicated geometries, a more accurate $\lambda(q,r)$ could depend on the resistance of the channels $r$, as well. In addition, if the properties of cells change with respect to time, $\lambda(q,r)$ could be a non-trivial function of time. we could further tune the clogging rate function to reflect the time-dependent or pressure-dependent physical properties of the cells based on experimental data or analytical models. For instance, in the case of sickling of the red blood cells, a clogging rate function that accurately captures the changes in deformability could lead to more accurate predictions about the failure time of the device. \par
In conclusion, studies on the effects of flow conditions and geometries on passing of deformable particles through narrow constrictions encounter a number of computational and analytical challenges. The challenges regarding the numerical procedures include the high computational cost and lack of accessibility of many of these resources. In addition, models that quantitatively describe how channel geometry and flow conditions influence particle deformation have not been fully developed \cite{def_nar_review}. This proves the value of a simple mathematical model that addresses clogging in such circumstances.
\vskip6pt

\enlargethispage{20pt}

\section*{Acknowledgments}
We acknowledge funding from NSF grant DMS-1913093.

\appendix
  \setcounter{equation}{0}  % reset counter 
\section{Parameter Estimation}
\label{app:alpha}
\renewcommand{\theequation}{A\arabic{equation}}
{Throughout most of this paper, we work with the normalized particle concentration $\alpha=1$ and normalized channel flow $\tilde{q}(t) = \frac{q(t)}{Q_0}$, where $Q_0$ is the initial total flow rate. The primary application considered in this paper concern red blood cells. Thus, w}hen estimating the order of magnitude of the concentration of particles per volume {(number density)} $\bar{\alpha}$, we may use the concentration of the red blood cells in a sample of blood, $4-6 \times 10^{15}$ cells per cubic-meter \cite{cell_conc}. 
In addition, we need an estimate for the flow rate in each channel. Given the pressure difference of 22.6 mm $\text{H}_\text{2}\text{O}$ \cite{Du1422} or about 222Pa, and an estimate of the resistance of each channel $r$ approximated as a rectangular pipe (see Table \ref{tab:variables}), we use the Hagen–Poiseuille equation to calculate the initial flow
\begin{equation}
   {q_0}(t) = \frac{1}{n}\frac{\Delta P}{r},
\end{equation}   
where $n$ indicates the number of columns. 
Given a range of 1 to 100 for the value of $n$ and 1 to 1000 for $m$, the order of magnitude of $q$ varies from $10^{-18}$ to $10^{-13}$ $m^3/s$. Therefore, the rate of arrival of cells into channels falls within the approximate range
{\begin{equation}
    0.001 s^{-1} < \bar{\alpha} q < 100 s^{-1}.
\end{equation}}
Thus, when working with the normalized flow rate $\Tilde{q}$, the corresponding normalized concentration falls in the range from 0.001 to 100 units. This justifies our use of $\alpha=1$.

\begin{center}
\begin{table}[]
\begin{tabular}{| c| c| c|  }
\hline
 \textbf{parameter} & \textbf{symbol} & \textbf{range}  \\ 
 \hline
 pressure difference across the device & $\Delta P$  & $100-220$Pa \\ 
 \hline
 total flow rate & Q &  $10^{-15}-10^{-10} m^3/s$ 
  \\  
 \hline
 channel flow rate & q & $10^{-18}-10^{-13} m^3/s$  \\  
 \hline
  clogging rate & $\lambda$ & $2.4 \times 10^{-4} - 0.5 s^{-1} $  \\  
 \hline
 number density & $\bar{\alpha}$ & $4-6 \times 10^{15} m^{-3} $  \\  
 \hline
 hydraulic resistance of the device & $R$ & $5 \times 10^{11} - 3 \times 10^{16} kg \cdot m^{-4}/s $  \\  
 \hline
  hydraulic resistance of the channels & $r$ & $4 \times 10^{15} kg \cdot m^{-4}/s  $  \\  
 \hline
dynamic viscosity & $\eta$ & $4 \times 10^{-3} Pa \cdot s  $ \cite{viscosity} \\  
 \hline
\end{tabular}
\caption{\label{tab:variables}The table contains a list of all the variables used in this paper and an estimate for their value or range of values in SI units for an application to a device similar to the one from \cite{Du1422} shown in Figure \ref{fig:schem}\textbf{A}.}
\end{table}
\end{center}

\section{Independent Channels}
\label{Ap:Ind_channels}
\renewcommand{\theequation}{B\arabic{equation}}
Theoretical results from the case of independent channels help us further understand the other more complicated cases. Reliability engineering \cite{reliability_fault_tol} and probability theory provide the tools for arriving at these results. In reliability engineering, the reliability {(or survival)} function, typically denoted by $R(t)$, a function of time, refers to the probability of survival of a component up to time $t$. As an example, for component $i$ with a constant hazard rate (clogging rate) $\lambda_i$, 
\begin{equation}
    R_i(t) = e^{-\lambda_i t}.
\end{equation}
The reliability function makes it easier to calculate the failure time of the entire system. When $m$ independent components with identical clogging rate $\lambda$ are in parallel, the reliability function for this set of parallel resistors $R_p(t)$ follows
\begin{equation}
    R_p(t) = 1 - \prod_{i = 1}^{m} \left( 1 - e^{-\lambda_i t} \right) = 1 -  \left( 1 - e^{-\lambda t} \right)^{m},
\end{equation}
and if $n$ such sets are in series, the reliability of the entire system $R_s(t)$ would be
\begin{equation}
    R_s(t) = \left(  R_p(t) \right)^n.
\end{equation}
This becomes extremely useful since now we can calculate the mean time to failure, or time of clogging $T_\text{clog}$ of the device:
\begin{equation}
    T_\text{clog} = \int_0^{\infty} R_s(t) dt.
    \label{Tclogir}
\end{equation}
We obtain the theoretical results for the case of independent channels in this paper by using $\eqref{Tclogir}$ and numerical integration.

\section{Stability in Multiple-Column Devices with a Constant Pressure Difference}
\label{Ap:stability_P}
\renewcommand{\theequation}{C\arabic{equation}}
Given a multiple-column device with $m$ rows and $n$ columns, a constant pressure gradient $\Delta P$, {channel resistance $r$, and initial total flow rate $Q_0 = \frac{m \Delta P}{n r}$} we can investigate the type of feedback that the system encounters after a change in the number of clogged channels in one column. 
\subsubsection{Example 1}
{W}e {explore} the stability of the symmetric solution to {\eqref{eq:mf-multiplecolumns}} in which $f_i(t) = f(t)$ for all $i$ in a constant pressure gradient device{. As an example, we consider} the clogging rate function $\lambda(q) = \alpha \tilde{q}$ where $\alpha $ is the normalized particle concentration and $\tilde{q} = q/Q_0$. Substituting this clogging rate in \eqref{eq:mf-multiplecolumns} for column $j$, we have
\begin{equation}
\label{eq:alphaq}
    \dot{f_j}(t) =  \alpha \tilde{q_j} (1-f_j(t)).
\end{equation}
Using \eqref{eq:q}, we can write \eqref{eq:alphaq} as
\begin{equation}
\begin{aligned}
    \dot{f_j}(t) & = \alpha  \frac{n}{m \sum_{i=1}^n  \frac{ 1-f_j(t) }{ 1-f_i(t)}} (1-f_j(t))\\
    & = \frac{\alpha n}{m}\left({\sum_{i=1}^n  \frac{ 1 }{ 1-f_i(t)}}\right)^{-1},
    \end{aligned}
\end{equation}
the right-hand side of which will be the same for all the other columns. Thus, if two arbitrary columns $k $ and $j$ start to deviate due to a stochastic jump, we see a neutral feedback,
\begin{equation}
    \frac{\ud}{\ud t}(f_j(t) - f_k(t)) = 0. 
\end{equation}
Therefore, there is no effective restoring force after deviations from the symmetric solution. \par

\begin{figure}[h]
    \centering
    \includegraphics[width = 12cm]{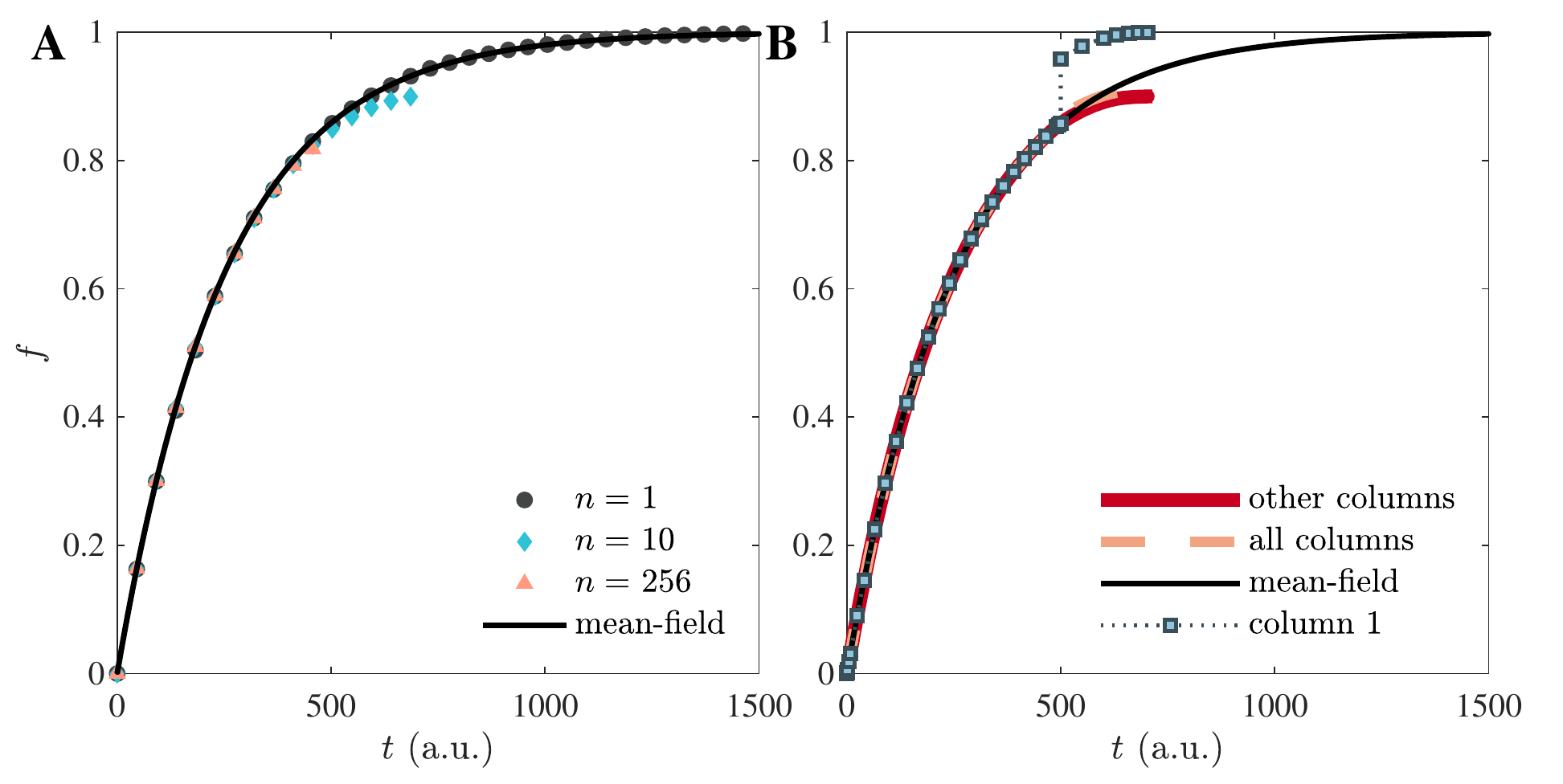}
    \caption{Stochastic jumps could explain why not all columns clog at the time of failure of a multiple-column device, here with fixed pressure.
\textbf{A} The stochastic simulations show that the fraction of clogged channels $f$ agrees with the mean-field prediction up to the clogging time of the device.  \textbf{B} Solving the system of differential equations numerically shows that when a set of columns, in this figure labeled column 1, makes a significant stochastic jump, {the magnitude of which was chosen arbitrarily here for demonstration purposes}, it expedites the clogging time of the device, and the fraction of clogged channels in other columns never reaches 1. {Note that the evolution of column 1 and other columns stops at around $t = 600$ corresponding to the time of clogging}. In these figures, the number of rows is $m = 256$ and the clogging rate function: $\lambda(\Tilde{q}) = \alpha \Tilde{q}$. The legend indicates the number of columns $n$ for each curve in \textbf{A}, and $n = 10$ in \textbf{B}.{The results in \textbf{A} show the mean of 100 trials.}     
}
    \label{fig:data_col_cp}
\end{figure}
Figure \ref{fig:data_col_cp} explores how the fraction of clogged channels in each column  $f_i$ deviates from the prediction of the mean-field model, in order to explain the results of Figures \ref{fig:cpf_T}\textbf{A} and \ref{fig:cpf_T}\textbf{B}. Figure \ref{fig:data_col_cp}\textbf{A} shows the results from the stochastic simulation for three different numbers of columns, as well as the mean-field model for a single-column device, all with the same number of rows, $m = 256$, and the same clogging rate function $\lambda(\Tilde{q}) = \alpha \Tilde{q}.$  As we see in these figures, even though the multiple-column devices fail before the predicted time of clogging of a single-column device, the mean fraction of clogged channels agrees with the mean-field prediction of a single-column device up to the time of clogging. Therefore, although the mean-field model cannot predict the time of clogging
in the case of multiple columns in the absence of additional information, it accurately predicts the fraction of clogged channels up to the time at which the device fails. \par
Figure \ref{fig:data_col_cp}\textbf{{B}} shows the numerical solution of a system of equations in which each equation corresponds to the fraction of clogged channels of each column. All columns start with the same initial condition and proceed in unison, as predicted by the mean-field model for a single column, until a stochastic jump occurs in one or more of the columns. We note that in practical applications, the stochastic effects break the symmetry; however, for the purpose of demonstration, the stochastic jump is externally imposed in this simulation to illustrate why the clogging time of multiple-column devices differs from single-column devices. In addition, it highlights how the device can fail before $f(t) = 1$. 
Investigating the stability of this symmetric solution provides insight into the behavior of the device. Depending on the conditions (constant pressure, constant flow, or independent channels) and the form of $\lambda(q)$, we see positive, negative, or neutral feedback after a stochastic jump. \par

\subsubsection{Example 2}
When $\lambda = \alpha \tilde{q}^2$, using \eqref{eq:q} and \eqref{eq:mf-multiplecolumns} and considering columns $j$ and $k$, we can write
\begin{equation}
\begin{aligned}
     \frac{\ud}{\ud t}\left(f_j(t) - f_k(t) \right) & = \alpha\left[  \frac{n}{m\sum_{i=1}^n  \frac{ 1-f_j(t) }{ 1-f_i(t)}} \right]^2 (1-f_j(t)) - \alpha\left[  \frac{n}{m\sum_{i=1}^n  \frac{ 1-f_k(t) }{ 1-f_i(t)}} \right]^2 (1-f_k(t)) \\
    & = \alpha \left[ \frac{n}{m\sum_{i=1}^n  \frac{ 1 }{ 1-f_i(t)}} \right]^2 [(1-f_j(t))^{-1} - (1-f_k(t))^{-1}] \\
    & = \alpha \left[ \frac{n}{m\sum_{i=1}^n  \frac{ 1 }{ 1-f_i(t)}} \right]^2 \frac{f_j(t) - f_k(t)}{(1-f_j(t))(1-f_k(t))}.
    \end{aligned}
\end{equation}
Thus, this is an example of a clogging rate function where we see positive feedback in a device with multiple columns and a constant pressure gradient. 
\subsubsection{Example 3: Independent Channels}
Given a device with $n$ columns with $m$ rows of independent channels all with a constant clogging rate $\lambda$, using \eqref{mf_eq} and for columns $j$ and $k$ we can write
\begin{equation}
    \frac{\ud}{\ud t}\left(f_j(t) - f_k(k) \right) = \lambda(1-f_j(t)) - \lambda(1-f_k(t)) = \lambda(f_k(t) - f_j(t)).
\end{equation}
So for the devices with independent channels with a constant clogging rate, we expect a negative feedback after stochastic jumps.
\section{Stability in Multiple-Column Devices with a Constant Total Flow Rate}
\renewcommand{\theequation}{D\arabic{equation}}
\label{Ap:stability_Q}

In a device with a total constant flow $Q_0$ and $\lambda(\Tilde{q}) = \alpha \tilde{q}(t)^\beta$, we may compare the rate of clogging of the columns $k $ and $j$ using \eqref{diff1}:
\begin{equation}
\label{eq:stabilityQ}
\begin{aligned}
    \frac{d}{dt}{(f_j(t) - f_k(t))} && =  \alpha \tilde{q_j}(t)^\beta ({1-f_j(t)}) - \alpha \tilde{q_k}(t)^\beta ({1-f_k(t)}) \\
    && = \alpha (\frac{1}{(1-f_j(t))^{\beta-1}} - \frac{1}{(1-f_k(t))^{\beta-1}}),
    \end{aligned}
\end{equation}
where we have used \eqref{eq:Qq} to obtain the final equality. When $\beta = 1$, the right-hand side of \eqref{eq:stabilityQ} goes to zero. Thus, this type of device has a neutral feedback, and differences between fraction of clogged channels in separate columns do not grow or shrink with time. On the other hand, when $\beta = 2$, we have a positive feedback: 
\begin{equation}
\begin{aligned}
    \frac{\ud}{\ud t}\left(f_j(t) - f_k(t)\right) = \alpha \frac{f_j(t) - f_k(t)}{(1-f_j(t))(1-f_k(t))} 
    \end{aligned}
\end{equation}
Thus, differences due to stochastic jumps grow larger over time, and we would expect a smaller fraction of clogged channels at the time of clogging.

\begin{figure}
    \centering
    \includegraphics[width = 12cm]{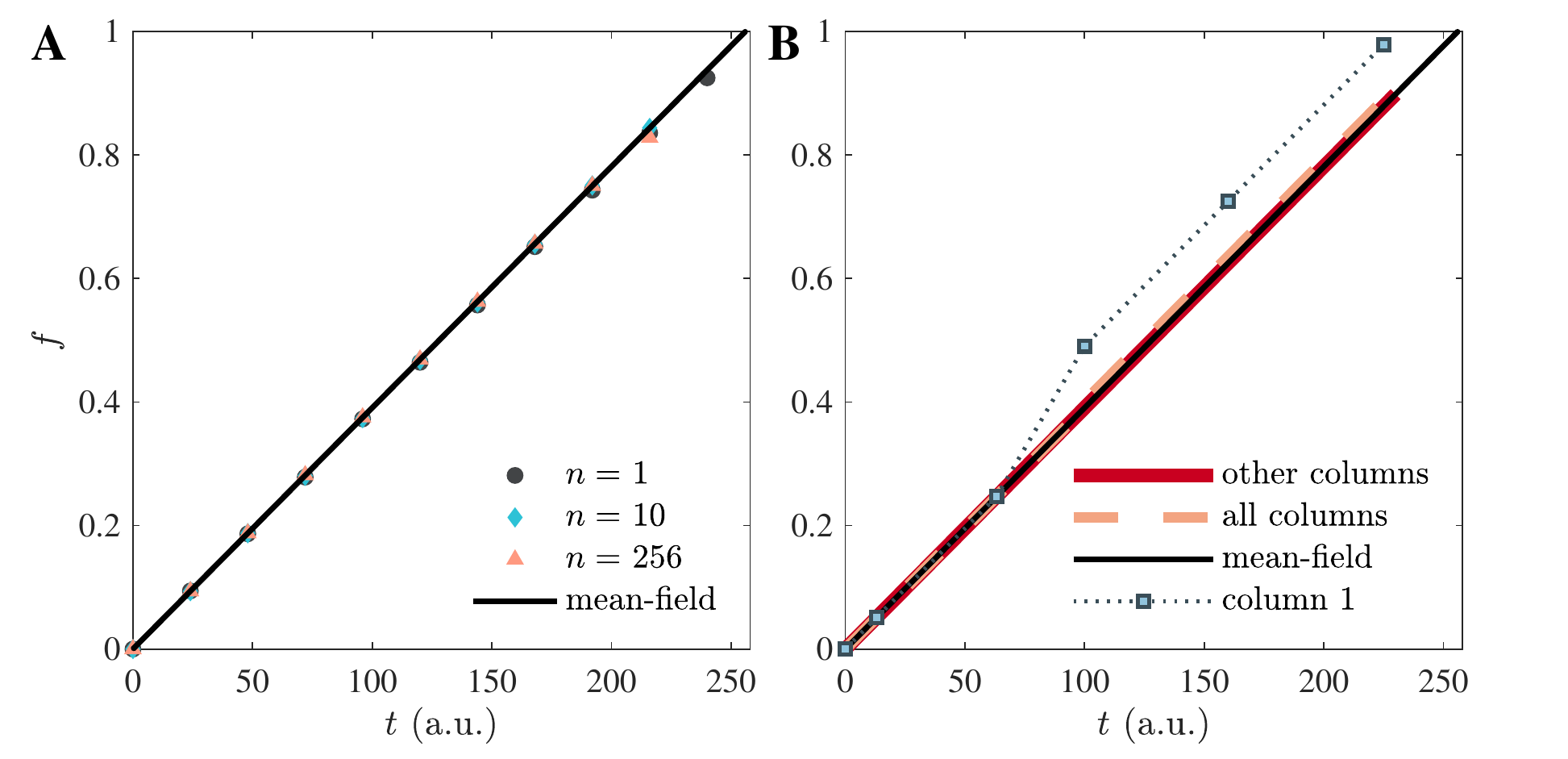}
    \caption{Stochastic jumps could explain why not all columns clog at the time of failure of a multiple-column device with $n$ columns, here with fixed total flow.
\textbf{A} The stochastic simulations show that the fraction of clogged channels $f(t)$ agrees with the mean-field prediction up to the clogging time of the device.  \textbf{B} Solving the system of differential equations numerically shows that when a set of columns, in this figure labeled column 1, makes a significant stochastic jump, {the magnitude and timing of which was picked arbitrarily here}, the fraction of clogged channels in the device never reaches 1 contrary to the mean-field prediction. In these figures, the number of rows is $m = 256$ and the clogging rate function: $\lambda(\Tilde{q}) = \alpha \Tilde{q}$. The legend indicates the number of columns for each curve in \textbf{A}, and $n = 10$ in \textbf{B}. {The results in \textbf{A} show the mean of 100 trials.}     }
    \label{fig:dat_col_cf}
\end{figure}

Figure \ref{fig:dat_col_cf}\textbf{A}  shows that the mean fraction of clogged channels from trials of stochastic simulations for different number of columns $n$ as a function of time. The mean-field curve illustrates the solution to \eqref{diff_i} for $m = 100$ and $\lambda(\Tilde{q}) = \alpha \Tilde{q} $. As in the case of constant pressure, for the constant flow case, the mean-field model correctly predicts the fraction of clogged channels up to the time at which the device fails. In addition, Figure \ref{fig:dat_col_cf}\textbf{B} demonstrates how stochastic effects may lead to a shorter lifetime for the device than the one predicted by the mean-field model. Similar to the case of constant pressure, in this example, the stochastic jump is externally imposed. 

%%%%%%%%%% Insert bibliography here %%%%%%%%%%%%%%
\vspace{0.5cm}
\bibliographystyle{unsrt}
\bibliography{microfluidics_gess_arxiv_v2}

\end{document}

%% file: File_Setup.tex
\usepackage{geometry}
\usepackage{caption}
\usepackage{epstopdf}
\usepackage{subcaption}
\usepackage[T1]{fontenc}
\usepackage[utf8]{inputenc}
\usepackage{amsmath, amsfonts, amsthm, amssymb} % Stærðfræðipakkar
\usepackage{braket, nicefrac} % fyrir mengi, brotabrot

\usepackage{siunitx}

\usepackage{enumitem, multicol}

\usepackage{graphicx, float} 
\usepackage{keystroke}
\usepackage{pgfplots}\usepgfplotslibrary{units}

\usepackage{tikz}
\usetikzlibrary{math}
\usepackage{fourier}
\usepackage[siunitx]{circuitikz} %%<---Circuit Diagrams

\usepackage{listings}
\usepackage{fancyvrb}

\definecolor{mygreen}{RGB}{28,172,0} 
\definecolor{mylilas}{RGB}{170,55,241}
\usepackage{tcolorbox}
\tcbuselibrary{skins}
\tcbset{matlabUT/.style={
		enhanced,
		colback=gray!1,
		colframe=gray!30,
		title=Command Window,
		arc=0mm,
		coltitle=black,
		center title, 
		title style={top color=white, bottom color = gray!30},
		grow to left by= -3 mm,
		left= 4 mm,
		grow to right by=0.5mm,
		colupper = gray!70!black
}}

\usepackage{hyperref}

\graphicspath{{graphics/}{Graphics/}{./}}

%% file: microfluidics_gess_arxiv_v2.bbl
\begin{thebibliography}{10}

\bibitem{icircuit1}
P.~Y. Paik, V.~K. Pamula, and K.~Chakrabarty.
\newblock Adaptive cooling of integrated circuits using digital microfluidics.
\newblock {\em IEEE Transactions on Very Large Scale Integration (VLSI)
  Systems}, 16(4):432--443, 2008.

\bibitem{SuWenjing2016FimA}
W.~Su, B.~S. Cook, Y.~Fang, and M.~M Tentzeris.
\newblock Fully inkjet-printed microfluidics: A solution to low-cost rapid
  three-dimensional microfluidics fabrication with numerous electrical and
  sensing applications.
\newblock {\em Scientific reports}, 6(1):35111, 2016.

\bibitem{RappBastianE2016MMMa}
B.~E. Rapp.
\newblock {\em Microfluidics: Modeling, Mechanics and Mathematics}.
\newblock Micro \& Nano Technologies Series. Elsevier, Oxford, United Kingdom,
  2016.

\bibitem{soil1}
M.~Hassanpourfard, R.~Ghosh, T.~Thundat, and A.~Kumar.
\newblock Dynamics of bacterial streamers induced clogging in microfluidic
  devices.
\newblock {\em Lab Chip}, 16(21):4091--4096, 2016.

\bibitem{soil2}
Filipe Felício, Vania Silverio, Sofia Duarte, Ana Galvão, Gabriel Monteiro,
  Susana Cardoso, and Rafaela Cardoso.
\newblock Preliminary tests on a microfluidic device to study pore clogging
  during biocementation.
\newblock {\em E3S Web of Conferences}, 92:11018, 2019.

\bibitem{canc_sort2}
Q.~Guo, S.~Duffy, K.~Matthews, E.~Islamzada, and H.~Ma.
\newblock Deformability based cell sorting using microfluidic ratchets enabling
  phenotypic separation of leukocytes directly from whole blood.
\newblock {\em Scientific Reports (Nature Publisher Group)}, 7:1--11, 2017.

\bibitem{culture_fouling}
L.~Yin, W.~Y. Au, C.~C. Yu, T.~Kwon, Z.~Lai, M.~Shang, M.~E. Warkiani,
  R.~Rosche, C.~T. Lim, and J.~Han.
\newblock Miniature auto‐perfusion bioreactor system with spiral microfluidic
  cell retention device.
\newblock {\em Biotechnology and bioengineering}, 118(5):1951--1961, 2021.

\bibitem{cl_fouling}
J.~R. Clapis, M.~J. Fan, and M.~L. Kovarik.
\newblock Supported bilayer membranes for reducing cell adhesion in
  microfluidic devices.
\newblock {\em Analytical methods}, 13(12):1535--154, 2021.

\bibitem{softmatter}
E.~Dressaire and A.~Sauret.
\newblock Clogging of microfluidic systems.
\newblock {\em Soft Matter}, 13:37--48, 2017.

\bibitem{canc_sort}
S.~Khetani, M.~Mohammadi, and A.~Sanati~Nezhad.
\newblock Filter-based isolation, enrichment, and characterization of
  circulating tumor cells.
\newblock {\em Biotechnology and Bioengineering}, 115:2504--2529, 2018.

\bibitem{Shelbymalaria_microfluidic}
J.~P. Shelby, J.~White, K.~Ganesan, P.~K. Rathod, and D.~T. Chiu.
\newblock A microfluidic model for single-cell capillary obstruction by
  plasmodium falciparum-infected erythrocytes.
\newblock {\em Proceedings of the National Academy of Sciences},
  100(25):14618--14622, 2003.

\bibitem{cancer1}
H.~W. Hou, Q.~S. Li, G.~Y.~H. Lee, A.~P. Kumar, C.~N. Ong, and C.~T. Lim.
\newblock Deformability study of breast cancer cells using microfluidics.
\newblock {\em Biomed Microdevices}, 11:557–564, 2009.

\bibitem{Du1422}
E.~Du, M.~Diez-Silva, G.~J. Kato, M.~Dao, and S.~Suresh.
\newblock Kinetics of sickle cell biorheology and implications for painful
  vasoocclusive crisis.
\newblock {\em Proceedings of the National Academy of Sciences},
  112(5):1422--1427, 2015.

\bibitem{WuTenghu2013malaria}
T.~Wu and J.~J. Feng.
\newblock Simulation of malaria-infected red blood cells in microfluidic
  channels: Passage and blockage.
\newblock {\em Biomicrofluidics}, 7(4):44115, 2013.

\bibitem{malaria2}
S.~M. Hosseini and J.~J. Feng.
\newblock How malaria parasites reduce the deformability of infected red blood
  cells.
\newblock {\em Biophysical Journal}, 103(1):1 -- 10, 2012.

\bibitem{BowHansen2011malaria}
H.~Bow, I.~V. Pivkin, M.~Diez-Silva, S.~J. Goldfless, M.~Dao, J.~C. Niles,
  S.~Suresh, and J.~Han.
\newblock A microfabricated deformability-based flow cytometer with application
  to malaria.
\newblock {\em Lab on a chip}, 11(6):1065--1073, 2011.

\bibitem{ManYuncheng2020OI}
Y.~Man, E.~Kucukal, R.~An, Q.~D Watson, J.~Bosch, P.~A. Zimmerman, J.~A.
  Little, and U.~A. Gurkan.
\newblock Microfluidic assessment of red blood cell mediated microvascular
  occlusion.
\newblock {\em Lab on a chip}, 2(12):286--299, 2020.

\bibitem{grav}
R.~Rowe and D.~Babcock.
\newblock Modelling the clogging of coarse gravel and tire shreds in column
  tests.
\newblock {\em Canadian Geotechnical Journal}, 44(11):1273--1285, 2007.

\bibitem{ped}
Y.~Tajima and T.~Nagatani.
\newblock Clogging transition of pedestrian flow in t-shaped channel.
\newblock {\em Physica A: Statistical Mechanics and its Applications},
  303(1-2):239--250, 2002.

\bibitem{ped_clg}
A.~Garcimartín, J.~M. Pastor, C.~Martín-Gómez, D.~Parisi, and I.~Zuriguel.
\newblock Pedestrian collective motion in competitive room evacuation.
\newblock {\em Scientific reports}, 7(1):10792--9, 2017.

\bibitem{SauretAlban2014Cbsi}
A.~Sauret, E.~C. Barney, A.~Perro, E.~Villermaux, H.~A. Stone, and
  E.~Dressaire.
\newblock Clogging by sieving in microchannels: Application to the detection of
  contaminants in colloidal suspensions.
\newblock {\em Applied physics letters}, 105(7):74101, 2014.

\bibitem{dis_sorrel}
S.~S. Massenburg.
\newblock {\em Clogging Mechanisms in Converging Microchannels}.
\newblock {PhD} dissertation, Harvard University, Graduate School of Arts \&
  Sciences, 2016.

\bibitem{sanaei2019}
P.~Sanaei and L.~J. Cummings.
\newblock Membrane filtration with multiple fouling mechanisms.
\newblock {\em Phys. Rev. Fluids}, 4:124301, 2019.

\bibitem{talbot_model}
C.~Barr\'e and J.~Talbot.
\newblock Cascading blockages in channel bundles.
\newblock {\em Phys. Rev. E}, 92:052141, 2015.

\bibitem{talbot_barre_reversible}
C.~Barr\'e, G.~Page, J.~Talbot, and P.~Viot.
\newblock Recurrence dynamics of particulate transport with reversible
  blockage: From a single channel to a bundle of coupled channels.
\newblock {\em Phys. Rev. E}, 99:042119, 2019.

\bibitem{SauretAlban2018Goci}
A.~Sauret, K.~Somszor, E.~Villermaux, and E.~Dressaire.
\newblock Growth of clogs in parallel microchannels.
\newblock {\em Physical review fluids}, 3(10), 2018.

\bibitem{connectivity}
P.~Bacchin, Q.~Derekx, D.~Veyret, K.~Glucina, and P.~Moulin.
\newblock Clogging of microporous channels networks: role of connectivity and
  tortuosity.
\newblock {\em Microfluidics and Nanofluidics}, 17(1):85--96, 2013.

\bibitem{droplet}
M.~Schindler and A.~Ajdari.
\newblock Droplet traffic in microfluidic networks: A simple model for
  understanding and designing.
\newblock {\em Physical review letters}, 100(4):044501--044501, 2008.

\bibitem{zebrafish}
S.~Chang, S.~Tu, K.~Baek, A.~Pietersen, Y.~Liu, V.~Savage, S.~Hwang, T.~Hsiai,
  and M.~Roper.
\newblock Optimal occlusion uniformly partitions red blood cells fluxes within
  a microvascular network.
\newblock {\em PLOS Computational Biology}, 13:12, 2017.

\bibitem{wyss}
H.~Wyss, D.~Blair, J.~Morris, H.~Stone, and D.~Weitz.
\newblock Mechanism for clogging of microchannels.
\newblock {\em Physical Review E}, 74:6, 2006.

\bibitem{Bruus}
H.~Bruus.
\newblock {\em Theoretical microfluidics}.
\newblock Oxford master series in physics ; 18. Oxford University Press,
  Oxford, United Kingdom, 2007.

\bibitem{Stone2007}
H.~A. Stone.
\newblock {\em Introduction to Fluid Dynamics for Microfluidic Flows}, pages
  5--30.
\newblock Springer US, Boston, MA, 2007.

\bibitem{AjdariArmand2004Sfin}
A.~Ajdari.
\newblock Steady flows in networks of microfluidic channels: building on the
  analogy with electrical circuits.
\newblock {\em C. R. Physique}, 5(5):539--546, 2004.

\bibitem{alim}
K.~Alim, S.~Parsa, D.~Weitz, and M.~Brenner.
\newblock Local pore size correlations determine flow distributions in porous
  media.
\newblock {\em Physical Review Letters}, 119:14, 2017.

\bibitem{AldousDavidJ1999DaSM}
D.~J. Aldous.
\newblock Deterministic and stochastic models for coalescence (aggregation and
  coagulation): A review of the mean-field theory for probabilists.
\newblock {\em Bernoulli : official journal of the Bernoulli Society for
  Mathematical Statistics and Probability}, 5(1):3--48, 1999.

\bibitem{reliability_eng}
A.~Murty and V.~Naikan.
\newblock Reliability strength design through inverse distributions:
  exponential and weibull cases.
\newblock {\em Reliability Engineering \& System Safety}, 54(1):77--82, 1996.

\bibitem{blokus2020multistate}
A.~Blokus.
\newblock {\em Multistate System Reliability with Dependencies}.
\newblock Academic Press, Cambridge, Massachusetts, 2020.

\bibitem{kusters}
R.~Kusters, T.~van~der Heijden, B.~Kaoui, J.~Harting, and C.~Storm.
\newblock Forced transport of deformable containers through narrow
  constrictions.
\newblock {\em Phys. Rev. E}, 90:033006, 2014.

\bibitem{Bielinski_Deformbality}
C.~Bielinski, O.~Aouane, J.~Harting, and B.~Kaoui.
\newblock Squeezing multiple soft particles into a constriction: Transition to
  clogging.
\newblock {\em Phys. Rev. E}, 104:065101, 2021.

\bibitem{Lange2015_deform}
J.~Lange, J.~Steinwachs, T.~Kolb, L.~Lautscham, I.~Harder, G.~Whyte, and
  B.~Fabry.
\newblock Microconstriction arrays for high-throughput quantitative
  measurements of cell mechanical properties.
\newblock {\em Biophysical journal}, 109(1):26--34, 2015.

\bibitem{sickling_pressure}
P.~Abbyad, P.~Tharaux, J.~Martin, C.~N. Baroud, and A.~Alexandrou.
\newblock Sickling of red blood cells through rapid oxygen exchange in
  microfluidic drops.
\newblock {\em Lab on a chip}, 10(19):2505--2512, 2010.

\bibitem{fbm1}
H.~E. Daniels.
\newblock The statistical theory of the strength of bundles of threads. i.
\newblock {\em Proceedings of the Royal Society of London. Series A,
  Mathematical and Physical Sciences}, 183(995):405--435, 1945.

\bibitem{JamesonGIncGamma}
G.~J.~O Jameson.
\newblock The incomplete gamma functions.
\newblock {\em Mathematical gazette}, 100(548):298--306, 2016.

\bibitem{cell_conc}
L.~Dean.
\newblock {\em Blood Groups and Red Cell Antigens}.
\newblock Bethesda (MD): National Center for Biotechnology Information, 2005.

\bibitem{perc_materials}
G.~Yakovlev, J.~D. Gran, D.~L. Turcotte, J.~B. Rundle, J.~R. Holliday, and
  W.~Klein.
\newblock A damage-mechanics model for fracture nucleation and propagation.
\newblock {\em Theoretical and Applied Fracture Mechanics}, 53(3):180--184,
  2010.

\bibitem{TsaiMichelle2012Ivmo}
M.~Tsai, A.~Kita, J.~L., R.~Rounsevell, J.~N. Huang, J.~Moake, R.~E. Ware,
  D.~A. Fletcher, and W.~A. Lam.
\newblock In vitro modeling of the microvascular occlusion and thrombosis that
  occur in hematologic diseases using microfluidic technology.
\newblock {\em The Journal of clinical investigation}, 122(1):408--418, 2012.

\bibitem{def_nar_review}
Z.~Zhang, J.~Xu, and C.~Drapaca.
\newblock Particle squeezing in narrow confinements.
\newblock {\em Microfluidics and nanofluidics}, 22(10):1--26, 2018.

\bibitem{viscosity}
E.~Nader, S.~Skinner, M.~Romana, R.~Fort, Nathalie Lemonne, Nicolas Guillot,
  Alexandra Gauthier, Sophie Antoine-Jonville, Céline Renoux, Marie-Dominique
  Hardy-Dessources, Emeric Stauffer, Philippe Joly, Yves Bertrand, and Philippe
  Connes.
\newblock Blood rheology: key parameters, impact on blood flow, role in sickle
  cell disease and effects of exercise.
\newblock {\em Frontiers in Physiology}, 10, 2019.

\bibitem{reliability_fault_tol}
B.~W. Johnson.
\newblock {\em Fault Tolerance}.
\newblock CRC Press LLC, 2000.

\end{thebibliography}
